\documentclass[12pt]{iopart}

\usepackage{subfig} 
\usepackage{graphicx} 

\newcommand{\mb}[1]{{\mathbf#1}} 

\begin{document}

\title[Gravitational Lensing by Black Holes in Astrophysics and in {\rm Interstellar}]{Gravitational Lensing by Spinning Black Holes in Astrophysics, and in the Movie \textit{Interstellar}}

\author{Oliver James$^1$,  Eug\'enie von Tunzelmann$^1$, Paul Franklin$^1$ and Kip S Thorne$^2$}

\address{$^1$Double Negative Ltd., 160 Great Portland Street, London W1W 5QA, UK}
\address{$^2$Walter Burke Institute for Theoretical Physics, California Institute of Technology, Pasadena, California 91125, USA}
\ead{oj@dneg.com, evt@dneg.com, paul@dneg.com, kip@caltech.edu}
\vspace{10pt}
\begin{indented}
\item[]Classical and Quantum Gravity {\bf 32} (2015) 065001. 
\item[]Received 27 November 2014, revised 12 January 2015 
\item[]Accepted for publication 13 January 2015
\item[]Published 13 February 2015 
\end{indented}

\begin{abstract}
\emph{Interstellar} is the first Hollywood movie to attempt depicting a black hole as it would actually be seen by somebody nearby.   For this, our team at \emph{Double Negative Visual Effects}, in collaboration with physicist Kip Thorne, developed a code called DNGR (Double Negative Gravitational Renderer) to solve the  equations for ray-bundle (light-beam) propagation through the curved spacetime of a spinning (Kerr) black hole, and to render IMAX-quality, rapidly changing images.    Our ray-bundle techniques  were crucial for achieving IMAX-quality smoothness without flickering; and they differ from physicists' image-generation techniques (which generally rely on individual light rays rather than ray bundles), and also differ from techniques previously used in the film
industry's CGI community.

This paper has four purposes:  (i) To describe DNGR for physicists and CGI practitioners, who may find interesting and useful some of our unconventional techniques. (ii) To present  the equations we use, when the camera is in arbitrary motion at an arbitrary location near a Kerr black hole, for mapping light sources to camera images via elliptical ray bundles.    (iii) To describe new insights,
from DNGR, into gravitational lensing when the camera is near the spinning 
black hole, rather than far away as in almost all prior studies; we focus 
on the shapes, sizes and influence of caustics and critical curves, the 
creation and annihilation of stellar images, the pattern of multiple images, and the influence of 
almost-trapped light rays, and we find similar results to the more familiar
case of a camera far from the hole. (iv) To describe how the images of the black hole Gargantua and its accretion disk, in the movie \emph{Interstellar}, were generated with DNGR---including, especially, the influences of (a) colour changes due to doppler and gravitational frequency shifts, (b) intensity changes due to the frequency shifts, (c) simulated camera lens flare, and (d) decisions that the film makers made about these influences and about the Gargantua's spin, with the goal of  producing images understandable for a mass audience.  There are no new astrophysical insights in this accretion-disk section of the paper, but disk novices may find it pedagogically interesting, and movie buffs may find its discussions of \emph{Interstellar} interesting.

\end{abstract}

\section{Introduction and overview}
\label{sec:Intro}

\subsection{Previous research and visualizations}
\label{subsec:Previous}

At a summer school in Les Houches France in summer 1972, James  Bardeen \cite{Bardeen73},  building on earlier work of  Brandon Carter \cite{Carter68}, initiated research on gravitational lensing by spinning black holes.  
Bardeen gave a thorough 
analytical analysis of null geodesics (light-ray propagation) around a spinning black hole; and,  
as part of his analysis, he computed how a black hole's spin affects the shape of
the shadow that the hole casts on light from a distant star field.  The shadow bulges out on the side
of the hole moving away from the observer, and squeezes inward and flattens 
on the side moving toward the observer.  The result, for a maximally spinning hole viewed from afar, is a D-shaped shadow;  cf. Figure \ref{Fig4:KerrLens} below. (When viewed up close, the shadow's flat edge has a shallow notch cut out of it, as hinted
by Figure \ref{Fig8:Nofingerprint} below.)

Despite this early work, gravitational lensing by black holes remained a backwater of physics research until decades later, when the prospect for actual observations brought it
to the fore.

There were, we think,  two especially memorable accomplishments in the backwater era.  The first was  a 1978 simulation of what a camera sees as it orbits a non-spinning black hole, with a star field in the background.  This simulation was carried out by  Leigh Palmer, Maurice Pryce and Bill Unruh \cite{Unruh78} on an Evans and Sutherland Vector
graphics display at Simon Fraser University.  Palmer, Pryce and Unruh did not publish their simulation, but
they showed a film clip from it in a number of lectures in that era.  The nicest modern-era film clip of this same sort that we know of is by  Alain Riazuelo (contained in his DVD \cite{RiazueloDVD} and available on the web at \cite{Riazuelo14}); see Figure \ref{Fig3:SchRays} and associated
discussion below.  And see \cite{MullerWeiskopf} for an online application
by Thomas M\"uller and Daniel Weiskopf for generating similar film clips.  Also of much interest in our modern
era are film clips by Andrew Hamilton \cite{Hamilton14} of what a camera sees when falling into
a nonspinning black hole; these have been shown at many planetariums, and elsewhere.

The other most memorable backwater-era accomplishment was a black and white simulation by Jean-Pierre Luminet \cite{Luminet78} of what a thin accretion disk, gravitationally lensed by a nonspinning black hole, would look like as seen from far away but close enough to resolve the image. In Figure \ref{Fig15:Disk}c below, we show a modern-era colour version of this, with the camera close to a fast-spinning black hole. 

Gravitational lensing by black holes began to be observationally important in the 1990s.  

Kevin Rauch and Roger Blandford \cite{Rauch94} recognised that, when a hot spot, in a black hole's accretion disk or jet, passes through caustics of the Earth's past light cone 
(caustics produced by the hole's spacetime curvature), the brightness of the hot spot's X-rays will 
undergo sharp oscillations with informative shapes.  This
has motivated a number of quantitative studies of the Kerr metric's caustics; see, especially  \cite{Rauch94,Bozza08,Bozza10} 
and references therein.

These papers' caustics are relevant for a source near the black hole and an observer far away, on Earth---in effect, on the black hole's ``celestial sphere'' at radius $r=\infty$.  In our paper, by contrast, we are interested in light sources that are usually on the
celestial sphere and an observer or camera
near the black hole.  For this reversed case, we shall discuss the relevant 
caustics in Sections \ref{subsec:KerrOuter} and \ref{subsec:Fingerprint}.   
This case has been
little studied, as it is of primarily cultural interest (``everyone" wants to know what it
would look like to live near a black hole, but nobody expects to make such 
observations in his or her lifetime), and of science-fiction interest.  Our paper
initiates the detailed study of this culturally interesting case; but we leave 
a full, systematic study to future research.  Most importantly, we keep our camera
outside the ergosphere --- by contrast with Alain Riazuelo's   \cite{Riazuelo12}
and Hamilton's \cite{Hamilton14} recent simulations with cameras deep inside the ergosphere and even plunging through the horizon.\footnote{We also have done
such simulations but not in enough detail to reveal much new.}

In the 1990s astrophysicists began to envision an era in which very long baseline
interferometry would make possible the imaging of black holes ---  specifically,
their shadows and their accretion disks.  This motivated visualizations, with ever
increasing sophistication, of accretion disks around black holes: modern variants
of Luminet's \cite{Luminet78} pioneering work.  See, especially, Fukue and Yokoyama \cite{Fukue88}, who
added colours to the disk; 
Viergutz \cite{Viergutz93}, who made his black hole spin, treated thick disks, and produced particularly nice and interesting coloured images and included the disk's secondary image which wraps under the
black hole; Marck \cite{Marck96}, who laid the foundations for a lovely movie now available on the web \cite{Marck91} with the camera moving around close to the disk, and who also included higher-order images, as did 
Fanton et.\ al.\ \cite{Fanton97} and Beckwith and Done \cite{Beckwith05}. 
See also papers cited in these articles.  

In the 2000s astrophysicists have focused on perfecting the mm-interferometer imaging
of black-hole shadows and disks, particularly the black hole at the centre of
our own Milky Way galaxy (Sgr A*).  See, e.g., the 2000 feasibility study by Falcke, Melia and Agol \cite{Falcke}.  See also references on the development and exploitation of GRMHD (general relativistic
magnetohydrodynamical) simulation codes for modelling accretion disks like
that in Sgr A* \cite{Gammie,Fragile,McKinney}; and references on detailed GRMHD models of Sgr A* and the models' comparison with observations  \cite{Moscibrodzka, Dexter, Broderick,BroderickGR}.  This is culminating in
a mm interferometric system called the \emph{Event Horizon Telescope} \cite{EHT}, which is beginning to yield interesting observational results though not
yet full images of the shadow and disk in Sgr A*.

All the astrophysical visualizations of gravitational lensing and accretion disks described above, and all others that we are aware of, are based on tracing huge numbers of light rays through curved spacetime.  A primary goal of today's
state-of-the-art, astrophysical ray-tracing codes (e.g., the Chan, Psaltis and \"Ozel's massively parallel, GPU-based code GRay \cite{GRay}) is very fast throughput, measured, e.g., in integration steps
per second; the spatial smoothness of images has been only a secondary concern.
For our \emph{Interstellar} work, by contrast, a primary goal is smoothness of
the images, so flickering is minimised when objects move rapidly across
an IMAX screen; fast throughput has been only a secondary concern.
  
With these 
different primary goals, in our own code, called DNGR, we have been driven to employ a different set of visualization
techniques from those of the astrophysics community---techniques based 
on propagation of ray bundles (light beams) instead of discrete light rays,
and on carefully designed spatial filtering to smooth the overlaps of neighbouring
beams; see Section \ref{sec:DNGR} and \ref{sec:App}.  Although, at Double Negative, we have
many GPU-based workstations, the bulk of our computational work
is done on a large compute cluster (the Double Negative render-farm)
that does not employ GPUs.  

In \ref{App:CompareCodes} we
shall give additional comparisons of our DNGR code with astrophysical codes and with other film-industry CGI codes.
 
\subsection{This paper}
\label{subsec:ThisPaper}

Our  work on gravitational lensing by black holes began in May 2013, when Christopher Nolan asked us to collaborate on building realistic images of a spinning black hole and its disk, with IMAX resolution, for his science fiction movie 
{\it Interstellar}.  We saw this  not only as an opportunity to bring realistic black holes
into the Hollywood arena, but also an opportunity to create a simulation code capable
of exploring a black hole's lensing with a level of image smoothness and dynamics not
previously available.

To achieve IMAX quality (with 23 million pixels per image and adequately smooth transitions between pixels), our code needed to integrate not only
rays (photon trajectories) from the light source to the simulated camera, but also bundles of  rays (light beams) with filtering to smooth the beams' overlap; see Section \ref{sec:DNGR} ,  \ref{subsec:App-RayBundle}, and \ref{subsec:App-Filtering}. 
And because the camera would sometimes be moving with speeds that are a substantial 
fraction of the speed of light, our code needed to incorporate relativistic aberration as well as Doppler shifts and gravitational redshifts.

Thorne, having had a bit of experience with this kind of stuff, put together a step-by-step prescription for how to map a light ray and ray bundle from the light source (the celestial sphere or 
an accretion disk) to the camera's local sky; see \ref{subsec:App-RayTrace} and \ref{subsec:App-RayBundle}.  He implemented
his prescription in Mathematica to be sure it produced images in accord with others'
prior simulations and his own intuition.  He then turned his prescription over to our Double Negative team, who created the fast, high-resolution code DNGR that we describe in
Section \ref{sec:DNGR} and \ref{sec:App}, and created the images to be lensed:  fields of stars and in some cases
also dust clouds, nebulae, and the accretion disk around {\it Interstellar}'s black hole, Gargantua.

In Section \ref{sec:StarField}, we use this code to explore a few interesting aspects of the 
lensing of a star field as viewed by a camera that moves around a fast-spinning black hole (0.999 of maximal---a value beyond which natural spin-down processes
become strong \cite{KipSpinDown}).  
We compute the intersection of the celestial sphere $(r=\infty$) with the primary, secondary, and tertiary caustics of the camera's past light cone and explore how the sizes and shapes of the resulting caustic curves change as the camera moves closer to the hole and as we go from the primary
caustic to higher-order caustics.  We compute the images of the first three caustics on the camera's
local sky (the first three critical curves), and we explore in detail the creation
and annihilation of stellar-image pairs, on the secondary critical curve, when a star passes 
twice through the secondary caustic.  We examine the tracks of stellar images on
the camera's local sky as the camera orbits the black hole in its equatorial plane.  
And we discover and explain why, as seen by a camera orbiting a fast spinning
black hole in
our galaxy, there is just one image of the galactic plane between adjacent critical
curves near the poles of the black hole's shadow, but there are multiple images between critical curves
near the flat, equatorial shadow edge.  The key to this from one viewpoint is the fact that higher-order caustics wrap around the celestial sphere multiple times, and from another viewpoint the key is light rays temporarily trapped
in prograde, almost circular orbits around the black hole.  By placing a checkerboard of paint swatches on the celestial sphere, we  explore in detail the overall gravitational lensing patterns seen by a camera near a fast-spinning black hole and the influence of aberration due to the camera's motion.

Whereas most of Section \ref{sec:StarField} on lensing of stellar images is new, 
in the context of a camera near the hole and stars far away, our Section \ref{sec:Disk},
on lensing of an accretion disk, retreads largely known ground.  But it does so in
the context of the movie {\it Interstellar}, to help readers understand the movie's 
black-hole images,  and does so in a manner that may be pedagogically interesting.
We begin with a picture of a gravitationally lensed disk made of equally spaced paint swatches.  This picture is useful for understanding the multiple images of the
disk that wrap over and under and in front of the hole's shadow.  We then replace the paint-swatch disk by a fairly realistic and rather thin disk (though one constructed by Double 
Negative artists instead of by solving astrophysicists' equations for thin accretion disks
\cite{NovikovThorne}).  We first compute the lensing of our semi-realistic disk but ignore  Doppler shifts and gravitational redshifts, which
we then turn on pedagogically in two steps:  the colour shifts and then the intensity shifts.   We discuss why Christopher Nolan and Paul Franklin chose to omit the Doppler shifts in the movie, and chose to slow the black hole's spin from
that, $a/M \simeq 1$, required to explain the huge time losses in 
\emph{Interstellar}, to $a/M = 0.6$.  And we then discuss and add simulated lens flare (light scattering and diffraction in the camera's lenses) for an IMAX camera that observes the disk -- something an astrophysicist would not
want to do because it hides the physics of the disk and the lensed galaxy beyond it,
but that is standard in movies, so computer generated images will have 
continuity with images shot by real cameras.

Finally, in Section \ref{sec:Conclusion} we summarise and we point to our use of
DNGR to produce images of gravitational lensing by wormholes.  

Throughout we use geometrized units in which $G$ (Newton's gravitation constant) and
$c$ (the speed of light) are set to unity, and we use the MTW sign
conventions \cite{MTW}.

\section{DNGR: our Double Negative Gravitational Renderer code}
\label{sec:DNGR}

Our computer code for making images of what a camera would see in the vicinity of a black hole or wormhole is
called the Double Negative Gravitational Renderer, or DNGR --- which obviously can
also be interpreted as the Double Negative General Relativistic code.
\begin{figure}[b!]
\begin{center}
\includegraphics[width=0.75\columnwidth]{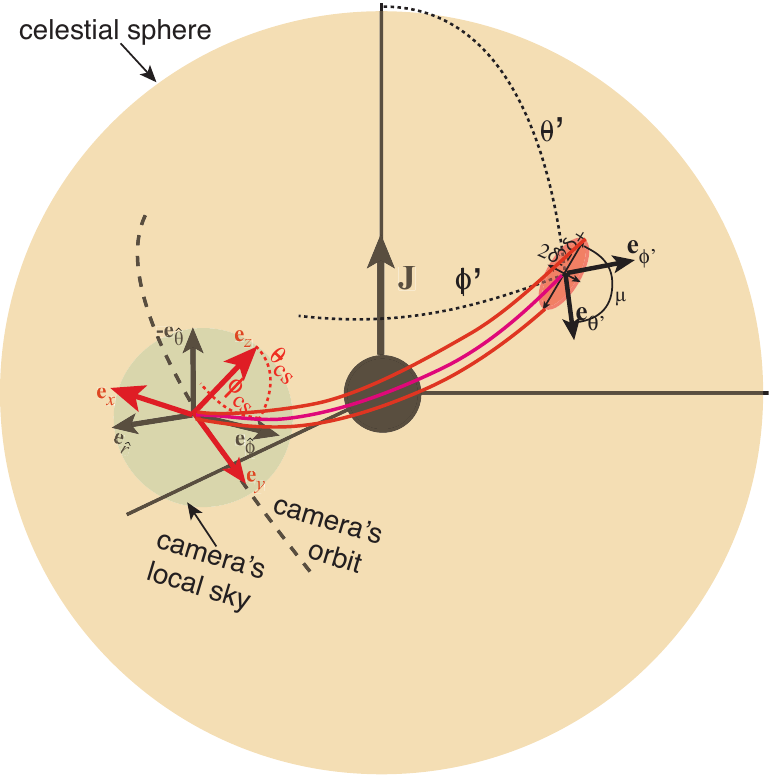}
\end{center}
\caption{The mapping of the camera's local sky $(\theta_{cs},\phi_{cs})$ onto the celestial sphere $(\theta',\phi')$ via a
backward directed light ray;  and the evolution of a ray bundle, that
is circular at the camera,  
backward along the ray to its origin, an ellipse on the celestial sphere.}
\label{Fig1:SkyMapA}
\end{figure}

\subsection{Ray Tracing}
\label{subsec:RayTracing}

The ray tracing part of DNGR produces a map from the celestial sphere (or the surface of an accretion disk) to the camera's local sky.  More specifically (see \ref{subsec:App-RayTrace}
for details and Figure \ref{Fig1:SkyMapA} for the ray-tracing geometry):

\begin{enumerate}
\item In DNGR, we adopt Boyer-Lindquist coordinates $(t,r,\theta,\phi)$ for the black hole's Kerr
spacetime.  At each event we introduce the locally non-rotating observer, also called
the fiducial observer or FIDO in the Membrane Paradigm \cite{MP}: the observer
whose 4-velocity is orthogonal to the surfaces of constant $t$, the Kerr metric's
space slices.  We regard the
FIDO as \emph{at rest} in space, and give the FIDO orthonormal basis vectors, 
$\{\mb e_{\hat r}, \mb e_{\hat \theta}, \mb e_{\hat \phi}\}$, that
point along the spatial coordinate lines. 
\item
We specify the camera's coordinate location $(r_c,\theta_c,\phi_c)$;
its direction of motion relative to the FIDO there, a unit vector $\bf B$ in the camera's reference
frame; and the camera's speed $\beta$ relative to the FIDO.
\item
In the camera's reference frame, we set up a right-handed set of three orthonormal basis 
vectors $\{\mb e_x, \mb e_y, \mb e_z\}$, 
with $\mb e_y \equiv \mb B$ along the direction of the camera's motion, $\mb e_x$
perpendicular to $\mb e_y$ and in the plane spanned by $\mb e_{\hat r}$ and 
$\e_{\hat \phi}$, and $\mb e_z$ orthogonal to $\mb e_x$ and $\mb e_y$.  See
Figure \ref{Fig1:SkyMapA}.  And we then set up a spherical polar coordinate system
$\{\theta_{cs}, \phi_{cs}\}$ for the camera's local sky (i.e.\ for the directions of
incoming light rays) in the usual manner dictated
by the camera's Cartesian basis vectors $\{\mb e_x, \mb e_y, \mb e_z\}$. 
\item
For a ray that  originates on
the celestial sphere (at $r=\infty$), we denote  the Boyer-Lindquist 
angular location at which it originates by $\{\theta',\phi'\}$.
\item
We integrate the null geodesic equation to propagate the ray from the camera to
the celestial sphere, thereby obtaining the map $\{ \theta'(\theta_{cs}, \phi_{cs} ),\phi'(\theta_{cs}, \phi_{cs} ) \}$ of points on the camera's local sky to points on the celestial sphere.
\item
If the ray originates on the surface of an accretion disk, we integrate the null geodesic
equation backward from the camera until it hits the disk's surface, and thereby 
deduce the map from a point on the disk's surface to one on the camera's sky.  For
more details on this case, see \ref{subsec:DNGRdisks}.
\item
We also compute, using the relevant Doppler shift and gravitational redshift, the
net frequency shift from the ray's source to the camera, and the corresponding net
change in light intensity.
\end{enumerate}

\subsection{Ray-Bundle (Light-Beam) Propagation}
\label{subsec:RayBundle}

DNGR achieves its IMAX-quality images by integrating a bundle of light rays (a light beam) backward
along the null geodesic from the camera to the celestial sphere 
using a slightly modified variant of a procedure
formulated in the 1970s by Serge Pineault and Robert Roeder \cite{PR1,PR2}.
This procedure is based on the equation of geodesic deviation and is
equivalent to the optical scalar equations \cite{Ehlers} that have
been widely used by astrophysicists in analytical (but not numerical) studies of gravitational lensing; see references
in Section 2.3 of \cite{Perlick}.
Our procedure, in brief outline, is this (see Figure \ref{Fig1:SkyMapA}); for full details,
see \ref{subsec:App-RayBundle}.
\begin{enumerate}
\item
In DNGR, we begin with an initially circular
(or sometimes initially elliptical) bundle of rays, with very small opening
angle, 
centred on a pixel on the camera's sky.
\item 
We  integrate the equation of geodesic
deviation backward in time along the bundle's central ray to deduce
the ellipse on the celestial sphere from which the ray bundle comes.  More 
specifically, we compute the angle $\mu$ that the ellipse's major axis makes with the
celestial sphere's $\mb e_{\theta'}$ direction, and the ellipse's major and
minor angular diameters $\delta_+$ and $\delta_-$ on the celestial sphere.  
\item
We then add up the spectrum and intensity of all the light emitted from within that ellipse;  and thence, using the
frequency and intensity shifts that were computed by ray tracing, we deduce the spectrum and intensity of the light arriving in the chosen camera pixel.
\end{enumerate}

\subsection{Filtering, Implementation, and Code Characteristics}
\label{subsec:DNGRDetails}

Novel types of filtering are key to generating our IMAX-quality images for movies.  In DNGR we use
spatial filtering to smooth the interfaces between beams (ray bundles), and temporal filtering to make dynamical images look like they were filmed with a movie camera.   For 
details, see   \ref{subsec:App-Filtering}.

In \ref{subsec:App-Implementation} we describe some details of our DNGR implementation of the ray-tracing, ray-bundle, and filtering equations; in 
\ref{subsec:App-CodeFarm} we describe some characteristics of our code and of 
Double Negative's Linux-based render-farm on which we do our computations; 
in \ref{subsec:DNGRdisks} we describe our DNGR modelling of accretion disks; and in 
\ref{App:CompareCodes} we briefly compare DNGR with other film-industry CGI codes and  state-of-the-art astrophysical simulation codes. 

\section{Lensing of a star field as seen by a moving camera near a black hole}
\label{sec:StarField}

\subsection{Nonspinning black hole}
\label{subsec:SchStarField}

In this subsection we review well known features of gravitational lensing by a nonspinning
(Schwarzschild) black hole, in preparation for discussing the same things for a
fast-spinning hole.

\begin{figure}[t!]
\begin{center}
\includegraphics[width=0.70\columnwidth]{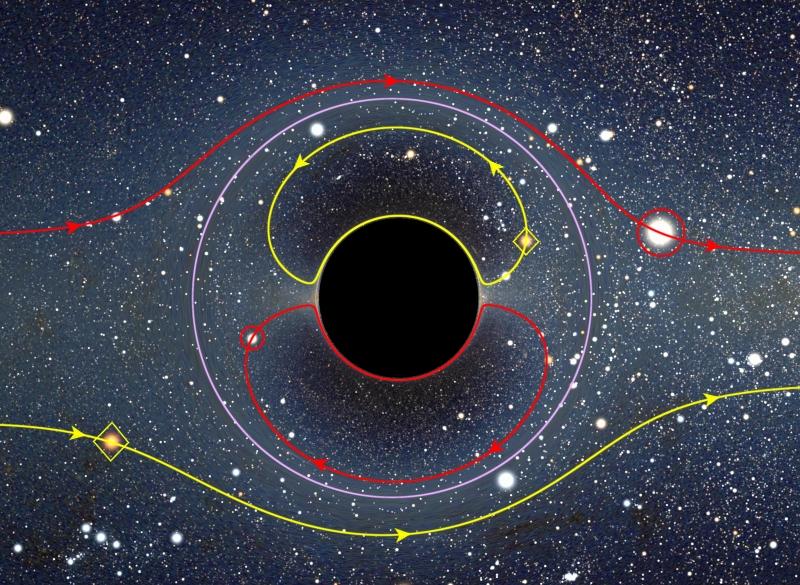}
\end{center}
\caption{Gravitational lensing of a star field by a nonspinning black hole, as seen by
a camera in a circular geodesic orbit at radius $r=30 M$.    
Picture courtesy Alain Riazuelo, from his film
clip  \cite{Riazuelo14}; coloured markings by us.
}
\label{Fig2:SchLens}
\end{figure}

We begin, pedagogically, with a still from a film clip by Alain Riazuelo \cite{Riazuelo14}, 
Figure \ref{Fig2:SchLens}.  The camera, at radius $r_c=30M$ (where $M$ is the 
black hole's mass) is moving in a circular geodesic orbit around the black hole, with a star field on the celestial sphere.  We focus on two stars,
which each produce two images on the camera's sky.  We place red circles around 
the images of the brighter star and yellow diamonds around those of the dimmer star.
As the camera orbits the hole, the images move around the camera's sky along the 
red and yellow curves.

Images outside the Einstein ring (the violet circle) move rightward and deflect away
from the ring.  These are called \emph{primary images}.  Images inside the Einstein ring (\emph{secondary images}) appear, in the film clip, to emerge from the edge
of the black hole's shadow, loop leftward around the hole, and descend back into
the shadow.  However, closer inspection with higher resolution reveals that their tracks actually close up along the shadow's edge as shown in the figure; the
close-up is not seen in the film clip because the images are so very dim along the inner leg of their tracks.   At all times, each
star's two images are on opposite sides of the shadow's centre.

This behaviour is generic.  Every star (if idealised as a point source of light), except a 
set of measure zero, has two images that behave in the same manner as the red and
yellow ones.  Outside the Einstein ring, the entire primary star field flows rightward, deflecting
around the ring; inside the ring, the entire secondary star field loops leftward, confined by the ring then back rightward along the shadow's edge. (There actually are more, unseen, images of the star field, even closer to the shadow's edge, that we shall 
discuss in Section \ref{subsec:KerrStarField}.)

\begin{figure}[t!]
\begin{center}
\includegraphics[width=1.00\columnwidth]{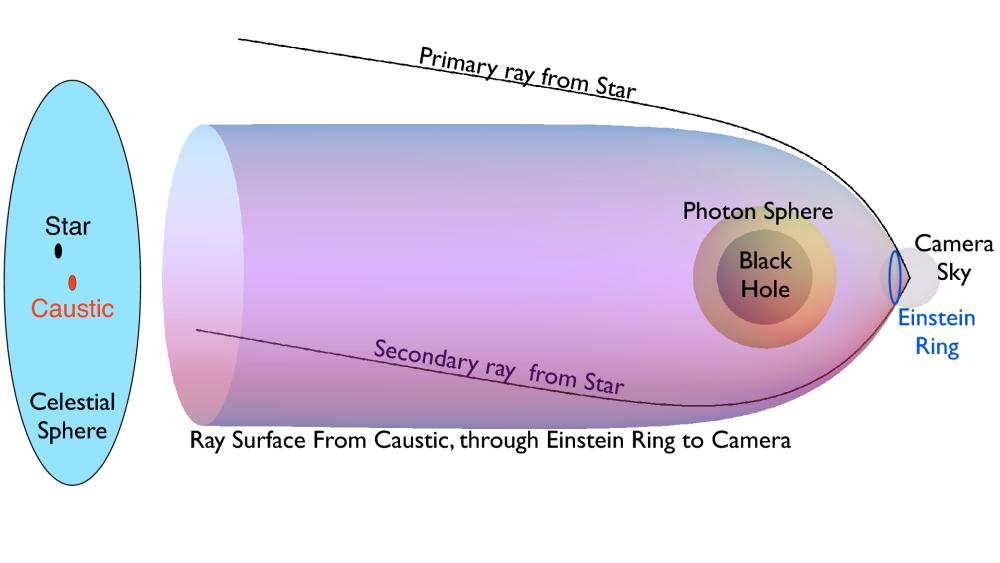}
\end{center}
\caption{Light rays around a Schwarzschild Black hole: geometric construction for explaining Figure \ref{Fig2:SchLens}. 
} 
\label{Fig3:SchRays}
\end{figure}

As is well known, this behaviour is easily understood by tracing light rays from the
camera to the celestial sphere; see Figure \ref{Fig3:SchRays}.

The Einstein ring is the image, on the camera's sky, of a point source that is on the celestial sphere, diametrically
opposite the camera; i.e., at the location indicated by the
red dot and labeled ``Caustic" in Figure \ref{Fig3:SchRays}.  Light rays from that
caustic point generate the purple ray surface that converges on the camera, and the
Einstein ring is the intersection of that ray surface with the camera's local sky.  

[The caustic point (red dot) is actually the intersection of the celestial 
sphere with a caustic line (a one-dimensional sharp edge) on the camera's past light cone.  This caustic line extends radially from the black hole's horizon
to the caustic point.]

The figure shows a single star (black dot) on the celestial sphere and two light rays
that travel from that star to the camera, gravitationally deflecting around
opposite sides of the black hole.  One of these rays, the primary one, arrives at the camera outside
the Einstein ring; the other, secondary ray, arrives inside the Einstein ring.  

Because the caustic point and the star on the celestial sphere both have dimension
zero,  as the camera moves, causing the caustic point to move relative to the star, there is zero probability
for it to pass through the star.  Therefore, the star's two images will never cross the Einstein
ring; one will remain forever outside it and the other inside---and similarly for all other
stars in the star field.  

However, if a star with finite size  passes close to the
ring, the gravitational lensing will momentarily stretch its two images  into lenticular shapes
that hug the Einstein ring and will produce a great, temporary increase in each image's energy flux at the camera
due to the temporary increase in the total solid angle subtended by each lenticular image.
This increase in flux still occurs when the star's actual size is too small for its images to
be resolved, and also in the limit of a point star.  For examples, see Riazuelo's film clip
\cite{Riazuelo14}.

(Large amplifications of extended images are actually seen in Nature, for example
in the gravitational lensing of distant galaxies by more nearby galaxies or galaxy
clusters; see, e.g., \cite{Bartelmann}.)

\subsection{Fast spinning black hole: introduction}
\label{subsec:KerrStarField}

\begin{figure}
\begin{center}
\includegraphics[width=1.0\columnwidth]{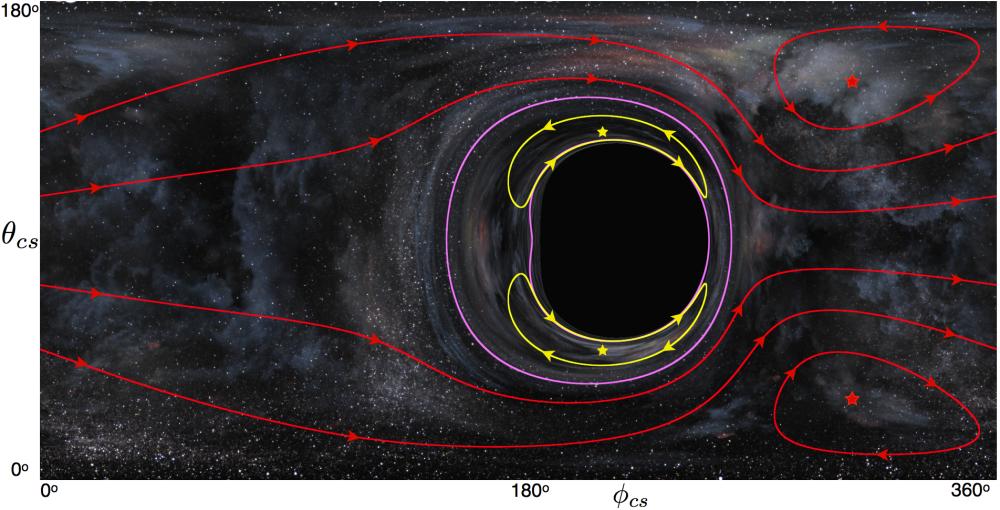}
\end{center}
\caption{Gravitational lensing of a star field by a  black hole with spin parameter
$a/M=0.999$, as seen by
a camera in a circular, equatorial geodesic orbit at radius $r_c=6.03 M$.
The red curves are the  trajectories of primary images, on the camera's sky, for stars at celestial-sphere latitudes $\theta'= 0.13\pi, 0.26\pi,  0.40\pi, 0.60\pi, 0.74\pi, 0.87\pi$.  The yellow curves are the 
trajectories of secondary images for stars at $\theta'=0.35 \pi, 0.65 \pi$.
The picture in this figure is a still from our first film clip
archived at \cite{DnegSite} and is copyright
\copyright 2015 Warner Bros. Entertainment Inc.
\emph{Interstellar} and all related characters and elements are
trademarks of and \copyright Warner Bros. Entertainment Inc. (s15).
The full figure appears in
the second and later printings of \emph{The Science of Interstellar}~\cite{TSI},
and is used by
permission of W. W. Norton \& Company, Inc. This image may be used under the terms
of the Creative Commons Attribution-NonCommercial-NoDerivs 3.0 (CC BY-NC-ND
3.0) license. Any further distribution of these images must maintain attribution to the
author(s) and the title of the work, journal citation and DOI. You may not use the
images for commercial purposes and if you remix, transform or build upon the images,
you may not distribute the modified images.
}
\label{Fig4:KerrLens}
\end{figure}

For a camera orbiting a  spinning black hole and a star field (plus sometimes dust
clouds and nebulae) on the celestial sphere, we have carried out a number of simulations with our code DNGR.  We show a few film clips from these simulations at \cite{DnegSite}.  Figure \ref{Fig4:KerrLens} is a still from one of
those film clips, in which the hole has spin $a/M=0.999$ (where $a$ is the hole's spin angular momentum per unit mass and $M$ is its mass), and the camera moves along a 
circular, equatorial, geodesic orbit at radius $r_c=6.03 M$.  

In the figure we show in violet two critical curves---analogs of the
Einstein ring for a nonspinning black hole.  These are images, on the
camera sky, of two caustic curves that reside on the celestial sphere;
see discussion below.  

We shall discuss in turn the region outside the secondary (inner) critical curve, and then
the region inside.

\subsection{Fast-Spinning Hole:  Outer region---outside the secondary critical curve}
\label{subsec:KerrOuter}

As the camera moves through one full orbit around the hole, the stellar images 
in the outer region make one full circuit along the red and yellow curves and other curves like them, 
largely avoiding the two critical curves of Figure \ref{Fig4:KerrLens}---particularly the outer (primary) one.

We denote by five-pointed symbols four images of two very special stars: stars that reside  where the hole's spin axis intersects the celestial sphere, $\theta'=0$ and $\pi$ ($0^{\rm o}$ and $180^{\rm o}$).  These are analogs of the Earth's star Polaris.  By symmetry, these pole-star images must remain fixed on the 
camera's sky as the camera moves along its circular equatorial orbit.  Outside
the \emph{primary (outer) critical curve}, all northern-hemisphere stellar images (images with $\theta_{\rm cs} < 90^{\rm o}$) circulate clockwise around the lower red pole-star image, and southern-hemisphere
stellar images, counterclockwise around the upper red pole-star image.  Between the primary and secondary (inner) critical curves,
the circulations are reversed, so at the primary critical curve there is a divergent shear in the image flow.

[For a nonspinning black hole (Figure \ref{Fig2:SchLens} above) there are also two critical curves, with the stellar-image motions confined by them: the Einstein ring, and a 
circular inner critical curve very close to the
black hole's shadow, that prevents the inner star tracks from plunging into the shadow and deflects them around the shadow so they close up.] 

\subsubsection{Primary and secondary critical curves and their caustics}
\label{subsec:KerrCaustics}

After seeing these stellar-image motions in our simulations, we explored the nature of the critical
curves and caustics for a camera near a fast-spinning black hole, and their influence.  Our exploration, conceptually, is a rather 
straightforward
generalisation of ideas laid out by Rauch and Blandford \cite{Rauch94} and by
Bozza \cite{Bozza08}.  They studied
a camera or observer on the celestial sphere and light sources orbiting a black hole;
our case is the inverse: a camera orbiting the hole and light sources on the celestial 
sphere.  

Just as the Einstein ring, for a nonspinning black hole, is the image of a caustic
point on the celestial sphere---the intersection of the celestial sphere with a caustic line on the camera's past light cone---so the critical curves for our spinning black
hole are also images of the intersection of the celestial sphere with light-cone caustics.  But the spinning hole's light-cone caustics generically are 2-dimensional surfaces
(folds) in the three-dimensional light cone,
so their intersections with the celestial sphere are one-dimensional: they are closed \emph{caustic curves} in the celestial sphere, rather than caustic points.  
The hole's rotation breaks spherical symmetry and converts non-generic caustic points into generic caustic curves.  (For this reason, theorems about caustics in the Schwarzschild spacetime, which are rather easy to prove, are of minor importance compared to results about generic caustics in the Kerr spacetime.)

We have computed the caustic curves, for a camera near a spinning black hole, by
propagating ray bundles backward from a fine grid of points on the camera's
local sky, and searching for points on the celestial sphere where the
arriving ray bundle's minor angular diameter $\delta_-$ passes through zero. 
All such
points lie on a caustic, and the locations on the camera sky where their ray 
bundles originate lie on critical curves. 

\begin{figure}[t!]
\begin{center}
\includegraphics[width=1.0\columnwidth]{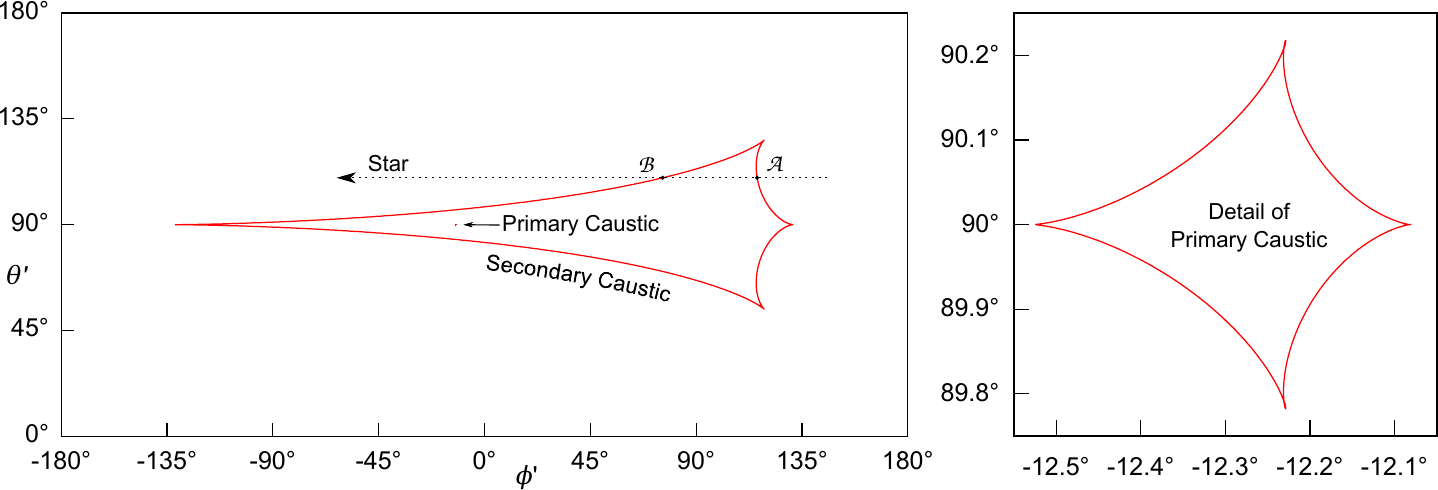}
\end{center}
\caption{The primary and secondary caustics on the celestial sphere, for the past light cone
of a camera moving along a circular, equatorial, geodesic orbit at radius $r_c= 6.03 M$ around a black hole with
spin parameter $a=0.999M$.  As the camera moves, in the camera's  
reference frame a star at $\theta'=0.608\pi$ travels along the dashed-line path.}
\label{Fig5:Astroid603}
\end{figure}

For a camera in the equatorial plane at radius $r_c=6.03 M$,
Figure \ref{Fig5:Astroid603} shows the primary and secondary caustic
curves.  These are images, in the celestial sphere, of the primary and 
secondary critical curves shown in Figure \ref{Fig4:KerrLens}. 

The primary caustic is a very small astroid (a four-sided figure whose sides
are fold caustics and meet in cusps).  It spans just $0.45^{\rm o}$  
in both the 
$\phi'$ and $\theta'$ directions.  The secondary caustic, by contrast, is a 
large astroid: it extends over $263^{\rm o}$ in $\phi'$ and $72^{\rm o}$ in 
$\theta'$.  All stars within
$36^{\rm o}$ of the equator encounter it as the camera, at $r_c=6.03M$,  orbits the black hole.  This is similar to the case of a source near the black hole and a camera on the celestial sphere (far from the hole, e.g.\ on Earth) \cite{Bozza08,Bozza10}.  There, also, the primary caustic is small and the secondary
large.  In both cases the dragging of inertial frames stretches the 
secondary caustic out in the $\phi$ direction.

 \subsubsection{Image creations and annihilations on critical curves}
  
Because the spinning hole's caustics have finite cross sections on the celestial
sphere, by contrast with 
the point caustics of a nonspinning black hole, stars, generically, can cross through them;
see, e.g., the dashed stellar path in Figure \ref{Fig5:Astroid603}.  As is well known from
the elementary theory of fold caustics (see, e.g., Section 7.5 of \cite{BT}), at each crossing two stellar images, on opposite sides of the caustic's critical curve,  merge and annihilate; or two are created.  And at the moment of creation or annihilation, the images are very bright. 

\begin{figure}[t!]
\begin{center}
\includegraphics[width=1.0\columnwidth]{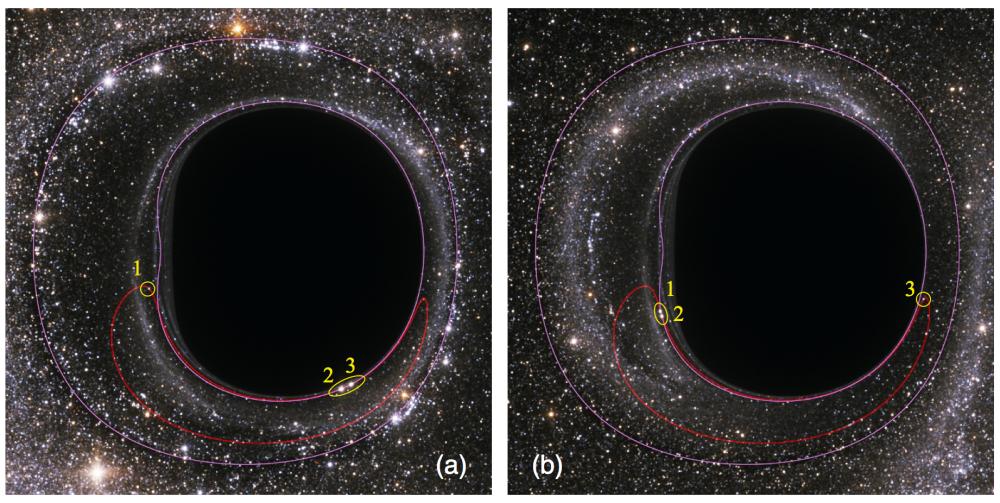}
\end{center}
\caption{Two stills from a film clip at \cite{DnegSite}).  In the left still, images 2 and 3 have just been created as their
star passed through caustic point $\mathcal A$ of Figure \ref{Fig5:Astroid603}a.
In the right still, images 1 and 2 are about to annihilate as their star passes
through caustic point $\mathcal B$.
}
\label{Fig6:CreateAnnihilate}
\end{figure}

Figure \ref{Fig6:CreateAnnihilate} (two stills from a film clip at \cite{DnegSite}) is an example of this. As the star
in Figure \ref{Fig5:Astroid603}a, at polar angle $\theta'=0.608\pi$, travels
around the celestial sphere relative to the camera, a subset of its stellar images 
 travels around the
red track of Figure \ref{Fig6:CreateAnnihilate}, just below the black hole's shadow.  (These are called the star's ``secondary
images'' because the light rays that bring them to the camera have the same number of poloidal turning points, one---and equatorial crossings, one---as the
light rays that map the secondary caustic onto the secondary critical curve; similarly these images' red track is called the star's ``secondary track''.)
At the moment of the left still, the star has just barely crossed the secondary caustic
at point $\mathcal A$ of Figure \ref{Fig5:Astroid603}a, and its two secondary stellar
images, \# 2 (inside the secondary critical curve) and \# 3 (outside it) have just been created at the point half way between \#2 and \#3,
where their red secondary track crosses the secondary critical curve (Figure \ref{Fig6:CreateAnnihilate}a). In the meantime,
stellar image \#1 is traveling slowly, alone, clockwise, around the track, outside the critical curve.
Between the left and right stills, image \#2 travels the track counter clockwise 
and images 1 and 3, clockwise. Immediately after the right still, the star
crosses the secondary caustic at point $\mathcal B$ in Figure \ref{Fig5:Astroid603}a, and the two images \#1 (outside the critical curve) and \#2 (inside it) annihilate at the intersection of their track
with the critical curve (Figure \ref{Fig6:CreateAnnihilate}b). As the
star in Figure \ref{Fig5:Astroid603}a continues on around the celestial sphere from point $\mathcal B$ to $\mathcal A$, the lone remaining image on the track,
image \#3, continues onward, clockwise, until it reaches the location \#1 of
the first still, and two new images are created at the location between \#2 and
\#3 of the first still; and so forth.   

\begin{figure}[t!]
\begin{center}
\includegraphics[width=0.95\columnwidth]{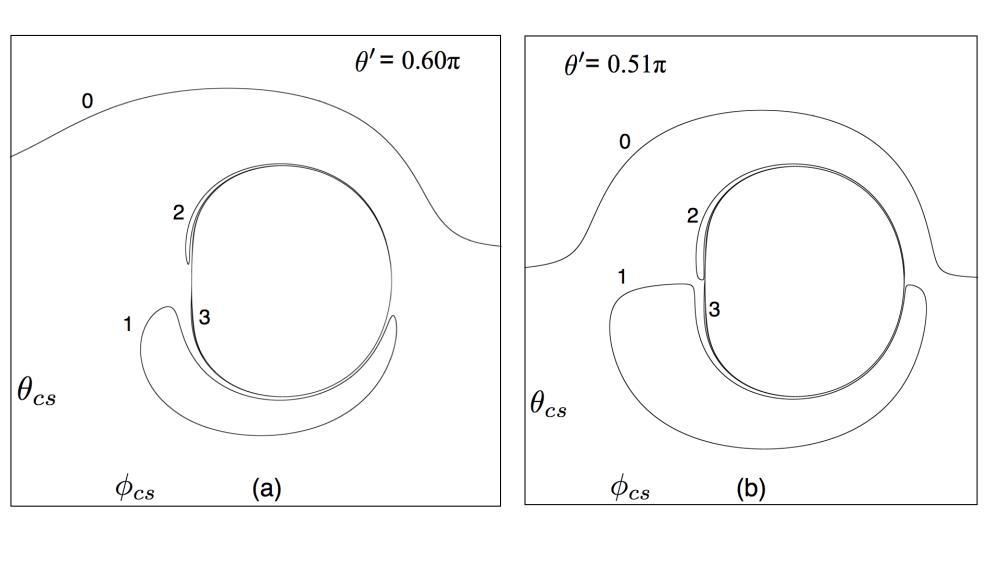}
\end{center}
\caption{ The tracks on the camera sky traveled by the multiple images of a single star, as the camera travels once around a black hole with $a/M=0.999$.  The camera's orbit
is a circular, equatorial geodesic with radius $r_c=6.03M$.  (a) For a star at
latitude $\theta'=0.60\pi$ ($18^{\rm o}$ above the equatorial plane; 
essentially the same star as in Figure \ref{Fig5:Astroid603}a).  (b) For a star at
$\theta'= 0.51 \pi$ ($1.8^{\rm o}$ above the equator).  The tracks are labeled
by the order of their stellar images (the number of poloidal, $\theta$, turning points on the
ray that brings an image to the camera).  
}
\label{Fig7:Latitude}
\end{figure}

Figure \ref{Fig7:Latitude}a puts these in a broader context. It shows the 
tracks of \emph{all} of the images of the star in Figure \ref{Fig5:Astroid603}a.  
Each image is labeled by its \emph{order} $n$: the number of poloidal $\theta$ turning points on the ray that travels to it from its star on the celestial sphere; or,
equally well (for our choice of a camera on the black hole's equator), the number of times that ray crosses 
the  equator $\theta=\pi/2$.  The order-0 track is called the primary track, and (with no ray-equator crossings) it lies on the same side of the equator as its star;
order-1 is the secondary track, and it lies on the opposite side of the equator from its star; order-2 is the tertiary track, on the same side of the equator as its star; etc.  

The primary track (order 0) does not intersect the primary critical curve, so a 
single primary image travels around it as the camera orbits the black hole.
The secondary track (order 1) is the one depicted red in 
Figure \ref{Fig6:CreateAnnihilate} and discussed above.  It crosses the secondary
critical curve twice, so there is a single pair creation event and a
single annihilation event; at some times there is a single secondary image on the track,
and at others there are three.  It is not clear to us whether the red secondary
track crosses the tertiary critical curve (not shown); but if it does, there will
be no pair creations or annihilations at the crossing points, because the
secondary track and the tertiary critical curve are generated by rays with different numbers of 
poloidal turning points, and so the  critical curve is incapable of 
influencing images on the track.  The extension to higher-order
tracks and critical curves, all closer to the hole's shadow, should be clear.
This pattern is qualitatively the same as when the light source is near the black hole and the camera far away, but in the hole's equatorial plane \cite{Bozza10}. 

And for stars at other latitudes the story is also the same; only the shapes of the
tracks are changed.  Figure \ref{Fig7:Latitude}b is an example. It shows
the tracks for a star just $1.8^{\rm o}$ above the black hole's equatorial
plane, at $\theta'=0.51\pi$.  

The film clips at \cite{DnegSite} exhibit these tracks and images all together, and show a plethora of image creations and annihilations.   Exploring these clips can be fun and informative.

\subsection{Fast-Spinning Hole:  Inner region---inside the secondary critical curve}
\label{subsec:Fingerprint}

The version of DNGR that we used for {\it Interstellar} showed a surprisingly complex,
fingerprint-like structure of gravitationally lensed stars inside the secondary critical curve, along the left side of the shadow.  

We searched for errors that might be responsible for it, and finding none, we thought it real. But Alain Riazuelo \cite{RiazueloNoFingerprint} saw nothing like it in his computed images. 
Making detailed comparisons with Riazuelo, we found a bug in
DNGR. When we corrected the bug, the complex pattern went
away, and we got excellent agreement with
Riazuelo (when using the same coordinate system), and with
images produced by Andy Bohn, Francois Hebert
and William Throwe using their Cornell/Caltech SXS imaging code \cite{SXS}.  Since the SXS code is so 
very different from ours (it is designed to visualize colliding black holes), that agreement
gives us high confidence in the results reported below.  

Fortunately, the bug we found had no noticeable impact on the images in \emph{Interstellar}.

With our debugged code, the inner region, inside the secondary critical curve, appears to be a continuation of
the pattern seen in the exterior region. 
There is a third critical curve within the second,
and there are signs of higher-order critical curves, all nested inside each other.  These are 
most visible near the flattened edge of the black hole's shadow on the side where the horizon's rotation is
toward the camera (the left side in this paper's figures).  The dragging of inertial frames
moves the critical curves outward from the shadow's flattened edge, enabling us to see things that otherwise could only be seen with a strong zoom-in.

\subsubsection{Critical curves and caustics for $r_c=2.60M$}

\begin{figure}[b!]
\begin{center}
\includegraphics[width=0.9\columnwidth]{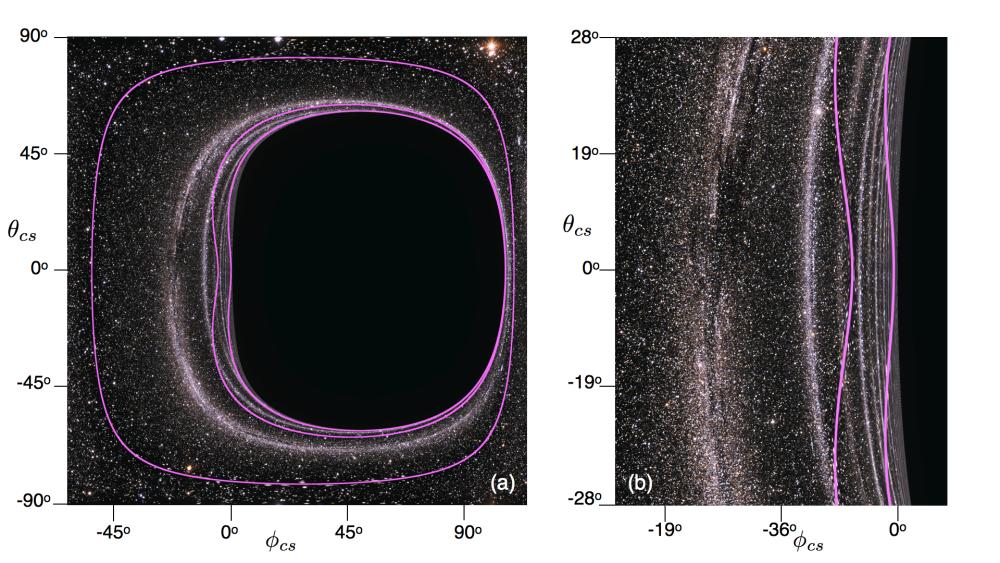}
\end{center}
\caption{(a) Black-hole shadow and three critical curves  for a camera traveling on a circular, geodesic, equatorial orbit at radius $r_c=2.60M$ in the equatorial plane of a black hole that has spin  $a/M=0.999$.   (b) Blowup of the equatorial region near the shadow's flat 
left edge.  The imaged star field is adapted from the Tycho-2 catalogue \cite{Tycho2} of the brightest 2.5 million stars seen from Earth, so it shows multiple images of the galactic plane.
}
\label{Fig8:Nofingerprint}
\end{figure}

\begin{figure}[h!]
\centering
\includegraphics[width=0.75\columnwidth]{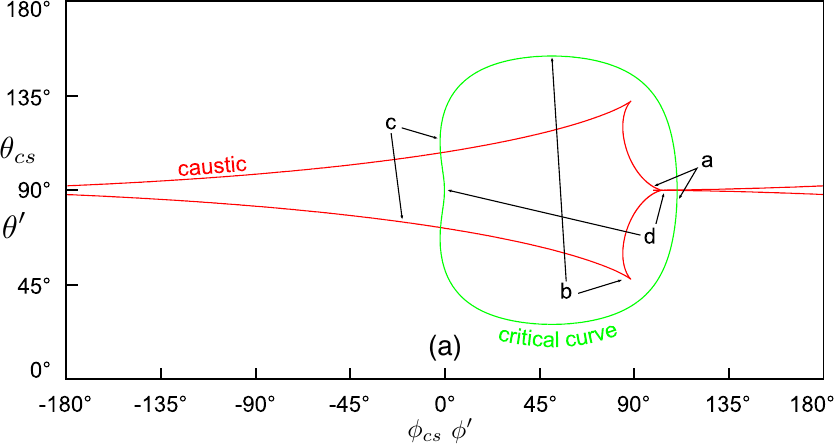}
\includegraphics[width=0.75\columnwidth]{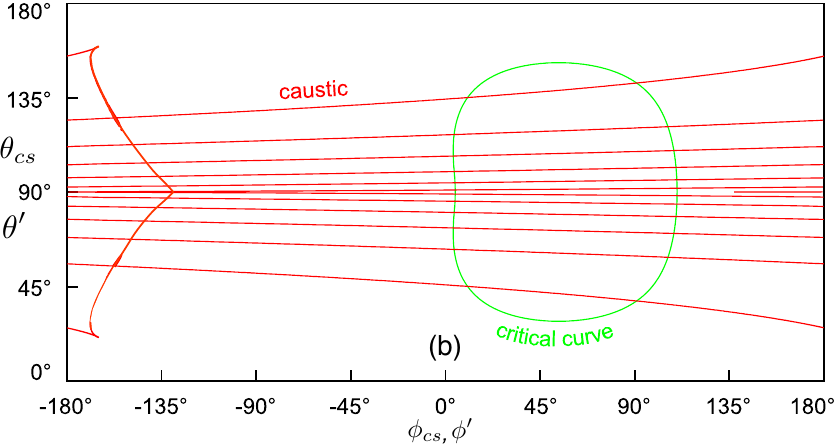}
\caption{(a) The secondary caustic (red) on the celestial sphere and secondary critical
curve (green) on the camera's sky, for the black hole and camera of Figure
\ref{Fig8:Nofingerprint}.  Points on each curve that are ray-mapped images of each other are marked by letters a, b, c, d.  (b) The tertiary caustic and tertiary critical curve.   
}
\label{Fig9:Astroid}
\end{figure}

The higher-order critical curves and the regions between them
can also be made more visible by moving the camera closer to the black hole's horizon.
In Figure \ref{Fig8:Nofingerprint} we have moved the camera in to $r_c=2.6M$ with $a/M=0.999$, and we show three nested critical 
curves. The primary caustic (the celestial-sphere image of the outer, primary, critical curve) is a tiny astroid, as at $r_c=6.03M$ (Figure \ref{Fig5:Astroid603}).  The secondary and tertiary
caustics are shown in Figure \ref{Fig9:Astroid}.  

By comparing these three figures with each other, we see that (i) 
as we move the camera closer to the horizon, the secondary caustics wrap 
further around the celestial sphere (this is reasonable since frame dragging
is stronger nearer the hole), and (ii) at fixed camera radius, 
each order caustic wraps further around the celestial sphere 
than the lower order ones.  
More specifically, for 
$r_c=2.60M$, the secondary caustic (Figure \ref{Fig9:Astroid}a) is stretched out to one full circuit around 
the celestial sphere compared to $2/3$ of a circuit when the camera is at $r_c=6.03M$ (Figure \ref{Fig5:Astroid603}),  and the tertiary caustic (Figure \ref{Fig9:Astroid}b) is stretched out 
to more than six circuits around the celestial sphere!  

The mapping of points between each caustic and its critical curve is displayed in film clips at \cite{DnegSite}.  For the secondary
caustic at $r_c=2.60M$, we show a few points in that mapping in Figure
\ref{Fig9:Astroid}a.  The left side of the critical curve (the
side where the horizon is moving toward the camera) maps into the long, 
stretched-out leftward sweep of the caustic.  The right side maps into the
caustic's two unstretched right scallops.     The same is true of other caustics and their critical curves.  

\subsubsection{Multiple images for $r_c=2.60M$}

Returning to the gravitationally lensed star-field image in Figure 
\ref{Fig8:Nofingerprint}b: notice the series of images of the galactic
plane (fuzzy white curves).  Above and below the black hole's shadow there is
just one galactic-plane image between the primary and secondary critical
curves, and just one between the secondary and tertiary critical curves.
This is what we expect from the example of a nonspinning black hole.
However, near the fast-spinning hole's left shadow edge, the pattern is
very different:  three galactic-plane images between the primary and 
secondary critical curves, and eight between the secondary and tertiary
critical curves. 

These multiple galactic-plane images are caused by the 
large sizes of the caustics---particularly their wrapping around the 
celestial sphere---and the resulting ease with which stars cross 
them, producing multiple stellar images.  An extension of an argument
by Bozza \cite{Bozza08} [paragraph preceding his Eq.\ (17)] makes this
more precise. (This argument will be highly plausible but not fully
rigorous because we have not developed a sufficiently complete
understanding to make it rigorous.)

Consider, for concreteness, a representative galactic-plane star that lies 
in the black hole's equatorial plane, inside the primary caustic, 
near the caustic's left cusp; and ask how many images the 
star produces on the camera's sky and where they lie. 
To answer this question, imagine moving the
star upward in $\theta'$ at fixed $\phi'$ until it is above all the celestial-sphere
caustics; then move it gradually back downward to its original, equatorial 
location.

When above the caustics, the star produces one image of each order: A 
primary image (no poloidal turning points) that presumably will remain
outside the primary critical curve when the star returns to its
equatorial location; a secondary image (one poloidal turning point) that
presumably will be between the primary and secondary critical curves
when the star returns; a tertiary image between the secondary and
tertiary critical curves; etc.

When the star moves downward through the upper left branch of the 
astroidal primary caustic, it creates two
primary images, one on each side of the primary critical curve.  When it
moves downward through the upper left branch of the secondary caustic (Figure
\ref{Fig9:Astroid}a),
it creates two secondary images, one on each side of the secondary
caustic.  And when it moves downward through the six sky-wrapped upper
left branches of the tertiary caustic (Figure \ref{Fig9:Astroid}b), it creates twelve tertiary images, 
six on each side of the tertiary caustic.  And because the upper left
branches of all three caustics map onto the left sides of their
corresponding critical curves, all the created images will wind up on
the left sides of the critical curves and thence the left side of the
black hole's shadow.  And by symmetry, with the camera and the star both
in the equatorial plane, all these images will wind up in the equatorial
plane.

So now we can count. In the equatorial plane to the left of the primary critical curve, there are
two images: one original primary image, and one caustic-created primary image.
These are to the left of the region depicted in  Figure \ref{Fig8:Nofingerprint}a.  Between the primary and
secondary critical curves there are three images: one original secondary image, 
one caustic-created primary, and one caustic-created secondary image.  These
are representative stellar images in the three galactic-plane images
between the primary and secondary critical curves  of Figure \ref{Fig8:Nofingerprint}.  And
between the secondary and tertiary critical curves there are eight stellar
images: one original tertiary, one caustic-created secondary, and six 
caustic-created tertiary images.  These are representative stellar images in
the eight galactic-plane images between the secondary and tertiary
critical curves of Figure \ref{Fig8:Nofingerprint}. 

This argument is not fully rigorous because: (i) We have not proved that
every caustic-branch crossing, from outside the astroid to inside,
creates an image pair rather than annihilating a pair; this is very
likely true, with annihilations occurring when a star moves out of the
astroid.  (ii) We have not proved that the original order-$n$ images wind
up at the claimed locations, between the order-$n$ and order-$(n+1)$
critical curves.  A more thorough study is needed to pin down these
issues.  
 
\begin{figure}[t!]
\begin{center}
\includegraphics[width=1.0\columnwidth]{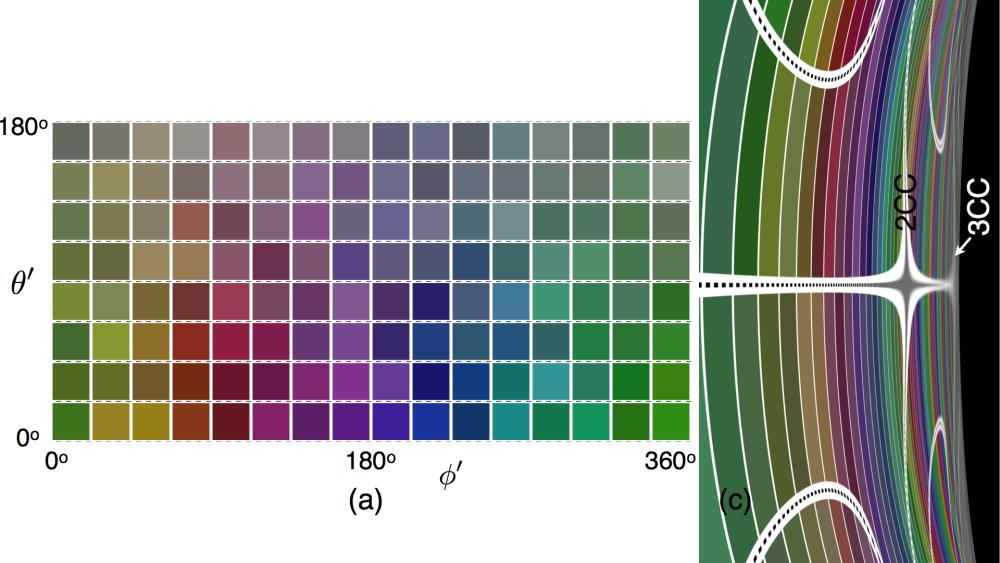}
\includegraphics[width=1.0\columnwidth]{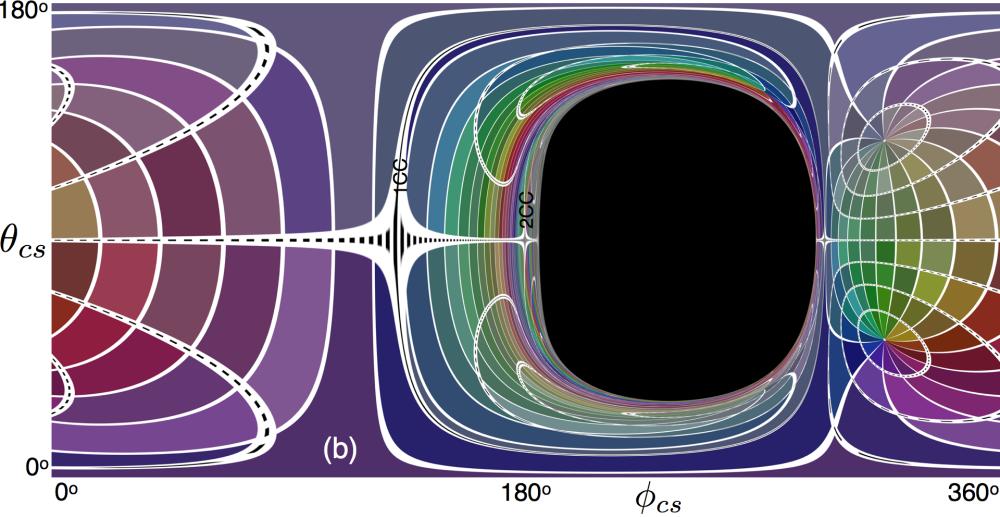}
\end{center}
\caption{(a) A checkerboard pattern of paint swatches placed on the celestial sphere 
of a black hole with
spin $a/M=0.999$.  As the camera, moves around a circular, equatorial, geodesic orbit 
at radius $r_c=2.60M$, stars 
move along horizontal dashed lines relative to the camera.  (b) This checkerboard 
pattern as seen gravitationally lensed on the camera's sky.  Stellar images move along the dashed 
curves.  The primary and secondary critical curves are labeled ``1CC" and ``2CC".  
(c) Blowup of the camera's sky near the left edge of the hole's shadow; ``3CC" is
the tertiary critical curve.}
\vskip2pc
\label{Fig10:Checker}
\end{figure}

\subsubsection{Checkerboard to elucidate the multiple-image pattern}

Figure \ref{Fig10:Checker} is designed to help readers explore this
multiple-image phenomenon in greater detail.  There we have placed, on
the celestial sphere, a checkerboard of paint swatches (Figure 
\ref{Fig10:Checker}a), with dashed lines running along the 
constant-latitude spaces between paint swatches, i.e., along the celestial-sphere
tracks of stars.  In Figure \ref{Fig10:Checker}b we show the gravitationally
lensed checkerboard on the camera's entire sky; and in Figure 
\ref{Fig10:Checker}c we show a blowup of the camera-sky region near the
left edge of the black hole's shadow.  We have labeled the critical curves
1CC, 2CC and 3CC for primary, secondary, and tertiary.  

The multiple images of lines of constant celestial-sphere longitude show up 
clearly in the blow-up, between pairs of critical curves; and the figure 
shows those lines being stretched vertically, enormously, in the vicinity of each
critical curve.  The dashed lines (star-image tracks) on the camera's sky 
show the same kind of pattern as we saw in Figure \ref{Fig7:Latitude}.  

\begin{figure}[t!]
\begin{center}
\includegraphics[width=0.75\columnwidth]{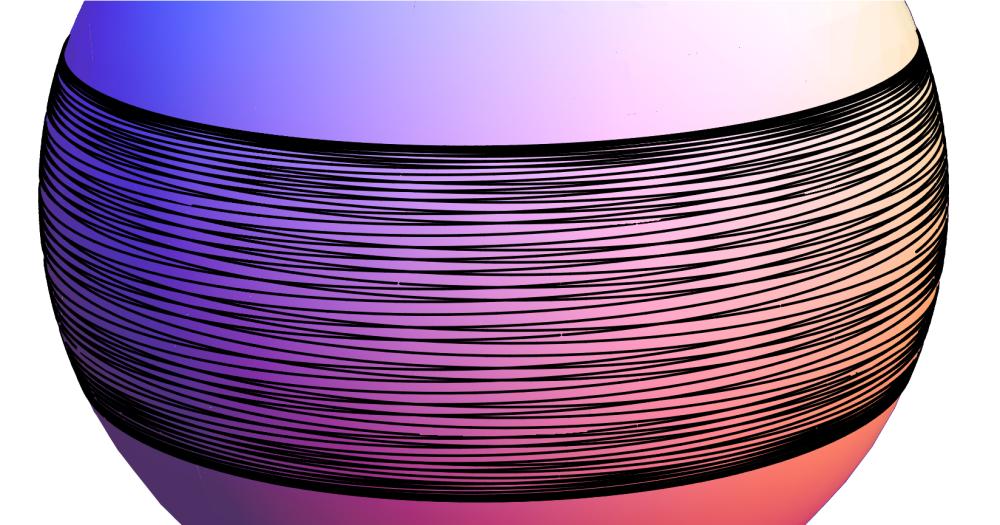}
\end{center}
\caption{A trapped, unstable, prograde photon orbit just outside the horizon of a 
black hole with spin $a=0.999 M$.  The orbit, which has constant Boyer-Lindquist coordinate radius $r$, is plotted on a sphere, treating its Boyler-Lindquist 
coordinates $(\theta,\phi)$ as though they were spherical polar coordinates.
}
\label{fig8:RingOfFire}
\end{figure}

\subsubsection{Multiple images explained by light-ray trapping}

The multiple images near the left edge of the shadow can also be understood 
in terms of the light rays that bring the stellar images to the camera.
Those light rays travel from the celestial sphere inward to near the black hole, where they get 
temporarily trapped, for a few round-trips, on near circular orbits
(orbits with nearly constant Boyer-Lindquist radius $r$), and then escape 
to the camera.  Each such nearly trapped ray is very close to
a truly (but unstably) trapped, constant-$r$ ray such as that shown
in Figure \ref{fig8:RingOfFire}.  These trapped rays (discussed in 
\cite{Teo} and in Chapters 6 and 8 of \cite{TSI}) wind up and down spherical
strips with very shallow pitch angles. 

As the camera makes each additional prograde trip around the black hole, the 
image carried by each temporarily trapped mapping ray gets wound around
the constant-$r$ sphere one more time (i.e., gets stored there for one
more circuit), and it comes out to the camera's sky slightly closer to
the shadow's edge and slightly higher or lower in latitude. Correspondingly,
as the camera moves, the star's image gradually sinks closer to the 
hole's shadow and gradually changes its latitude---actually moving away
from the equator when approaching a critical curve and toward the equator
when receding from a critical curve.  This behaviour is seen clearly,
near the shadow's left edge, in the film clips at 
\cite{DnegSite}.

\begin{figure}
\begin{center}
\includegraphics[width=0.75\columnwidth]{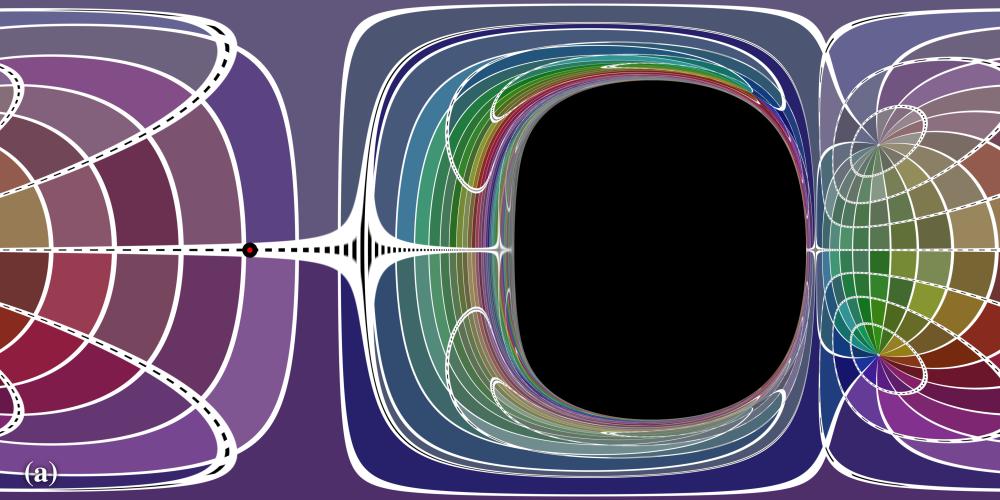}
\vskip0.5pc
\includegraphics[width=0.75\columnwidth]{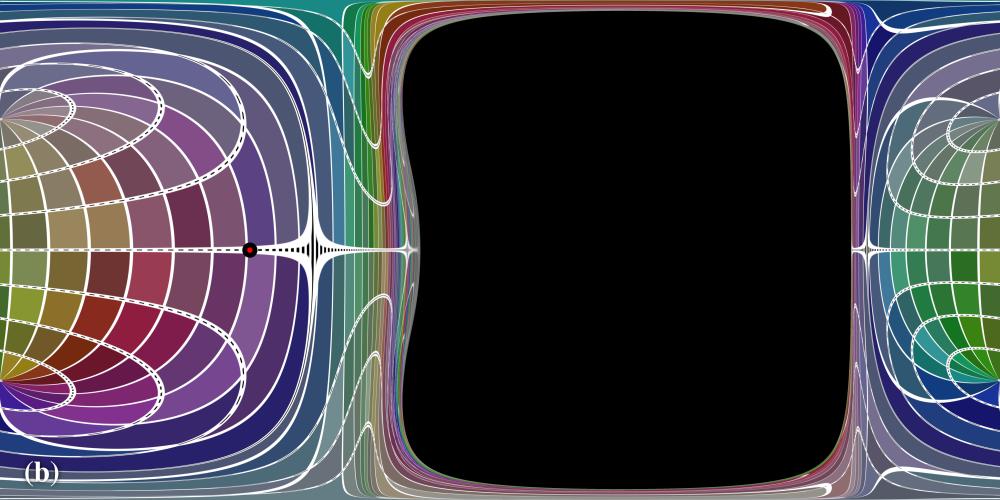}
\vskip0.5pc
\includegraphics[width=0.75\columnwidth]{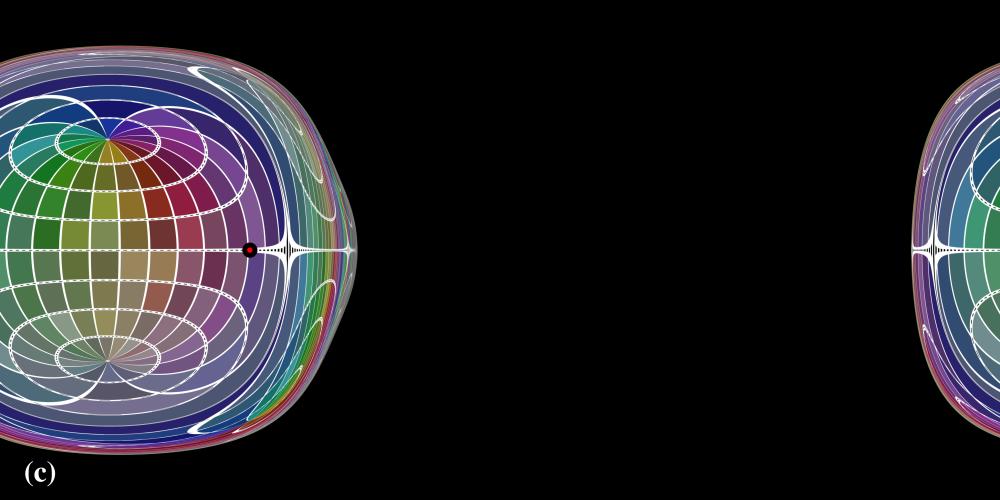}
\end{center}
\caption{Influence of aberration, due to camera motion, on gravitational lensing by a black hole with
$a/M=0.999$.  The celestial sphere is covered by the paint-swatch checkerboard of Figure 
\ref{Fig10:Checker}a, the camera is at radius $r_c=2.60M$ and is moving in the azimuthal ${\mb e}_{\hat \phi}$ direction, and the camera speed is: (a) that of a prograde, geodesic, circular orbit
(same as Figure \ref{Fig10:Checker}b),
(b) that of a zero-angular-momentum observer (a FIDO), and (c) at rest in the Boyer-Lindquist coordinate system.  The coordinates are the same as in Figure \ref{Fig10:Checker}b.  
}
\label{Fig12:Aberration}
\end{figure}

\subsection{Aberration: Influence of the camera's speed}

The gravitational lensing pattern is strongly influenced not only by the black hole's
spin and the camera's location, but also by the camera's orbital speed.  We explore
this in Figure \ref{Fig12:Aberration}, where we show the gravitationally lensed paint-swatch checkerboard of Figure \ref{Fig10:Checker}a for a black hole with spin $a/M=0.999$, a camera in the equatorial plane at radius $r_c=2.60M$, and three different camera velocities, 
all in the azimuthal ${\bf e}_{\hat \phi}$ direction: (a) camera moving along a prograde circular geodesic orbit
[coordinate angular velocity $d\phi/dt =1/(a + r_c^{3/2})$ in the notation of \ref{subsec:App-RayTrace}]; (b) camera moving along a zero-angular-momentum orbit 
[$d\phi/dt=  \omega$, Eq.\ (\ref{eq:KerrQuantities}), which is  speed $-0.546 c$ as measured by a circular, geodesic observer]; and (c) a static camera, i.e.\ at rest
in the Boyer-Lindquist coordinate system [$d\phi/dt = 0$, which is speed $-0.813 c$ as measured by
the circular, geodesic observer].

The huge differences in lensing pattern, for these three different camera velocities, are due, of course, to special relativistic aberration.  (We thank Alain Riazuelo for pointing out to us that  aberration effects should be huge.)  For prograde geodesic motion (top picture), the hole's shadow is relatively small and the sky around it, large.  As seen in the geodesic reference frame, the zero-angular-momentum camera and the static camera are moving in the direction of the red/black dot---i.e., toward the right part of the external universe and away from the right part of the black-hole shadow---at about half and 4/5 the speed of light respectively.  So 
the zero-angular-momentum camera (middle picture) sees the hole's shadow much enlarged due to aberration, and the external universe shrunken; and the static camera sees the shadow enlarged so much that it encompasses somewhat more than half the sky (more than $2\pi$ steradians), and sees the external universe correspondingly shrunk.

Despite these huge differences in lensing patterns, the multiplicity of images between critical curves is unchanged: still three images of some near-equator swatches between the primary and secondary
critical curves, and eight between the secondary and tertiary critical curves.  This is because the caustics in the camera's past light cone depend only on the camera's location and not on its velocity, so a point source's caustic crossings are independent of camera velocity, and the image pair 
creations and annihilations along critical curves are independent of camera velocity.

\section{Lensing of an accretion disk}
\label{sec:Disk}

\subsection{Effects of lensing, colour shift, and brightness shift: a pedagogical discussion}
\label{subsec:DiskPedagogy}

We have used our code, DNGR, to construct images of what a thin accretion disk in the equatorial plane of a fast-spinning black hole would
look like, seen up close.   For our own edification, we explored successively the
influence of the bending of light rays (gravitational lensing), the influence of Doppler frequency shifts and
gravitational frequency shifts on the disk's colours,  the influence of the  frequency shifts on the brightness of the disk's light, and the influence of 
\emph{lens flare} due to light scattering and diffraction in the lenses of a simulated
65 mm IMAX camera.  Although all these issues except lens flare have
been explored previously, e.g. in \cite{Luminet78,Fukue88,Viergutz93,Fanton97,Beckwith05} and references therein, our images may be of pedagogical interest, so we show them here.  We also show them as a foundation for discussing the choices that were made for \emph{Interstellar's}
accretion disk.

\begin{figure}[t!]
\begin{center}
\includegraphics[width=1.0\columnwidth]{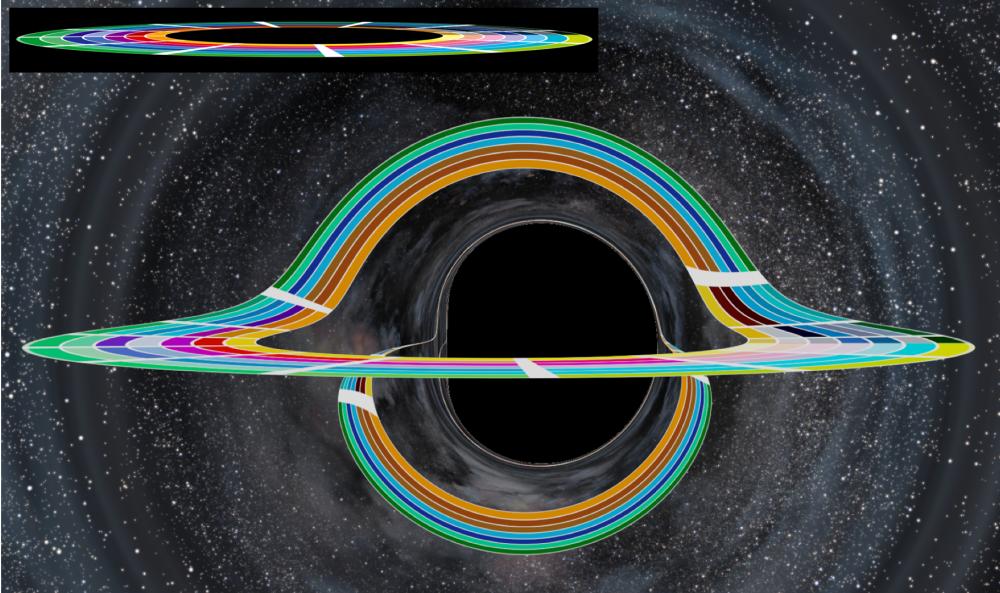}
\end{center}
\caption{Inset: Paint-swatch accretion disk with inner and outer radii
$r=9.26M$ and $r=18.70M$ before being placed around a black hole. 
Body: This paint-swatch disk, now in the equatorial plane around a black hole
with $a/M=0.999$, as viewed by a
camera   at $r_c = 74.1 M$ and $\theta_c = 1.511$ ($86.56^{\rm o}$), ignoring frequency shifts, associated colour and 
brightness changes, and lens flare.   (Figure from \emph{The Science of Interstellar} \cite{TSI}, 
used by permission of W. W. Norton \& Company, Inc,
and created by our Double Negative team, TM \& \copyright Warner Bros. Entertainment Inc. (s15)).
This image may be used under the terms
of the Creative Commons Attribution-NonCommercial-NoDerivs 3.0 (CC BY-NC-ND
3.0) license. Any further distribution of these images must maintain attribution to the
author(s) and the title of the work, journal citation and DOI. You may not use the
images for commercial purposes and if you remix, transform or build upon the images,
you may not distribute the modified images.
}
\label{Fig13:DiskPaintSwatch}
\end{figure}

\subsubsection{Gravitational lensing}
\label{subsec:DiskLensing}

Figure \ref{Fig13:DiskPaintSwatch} illustrates the influence of gravitational lensing (light-ray bending).
To construct this image, we placed the infinitely thin disk shown in the upper left
in the equatorial plane around a fast-spinning black hole, and we used DNGR to compute what the disk
would look like to a camera near the hole and slightly above the disk plane.  The disk consisted of paint swatches arranged in a simple pattern that
facilitates identifying, visually, the mapping from the disk to its lensed images. We omitted frequency shifts and their associated colour and brightness changes, and also omitted camera  
lens flare; i.e., we (incorrectly) transported the light's specific intensity
$
I_\nu = dE/dt\, dA\, d\Omega\, d\nu
$
along each ray, unchanged, as would be appropriate in flat spacetime.  Here $E$, $t$, $A$, $\Omega$, and $\nu$ are energy, time, area,
solid angle, and frequency measured by an observer just above the disk or at the
camera.

In the figure we see three  images of the disk.  The upper image swings around
the front of the black hole's shadow and then, instead of passing behind the shadow, it
swings up over the shadow and back down to close on itself.  This wrapping over the
shadow has a simple physical origin:  Light rays from the top face of the disk, which is
actually behind the hole, pass up over the top of the hole and down to the camera 
due to gravitational light deflection; see Figure 9.8 of \cite{TSI}.  This entire image comes
from light rays emitted by the disk's top face.  By looking at the colours, lengths, and widths of the disk's swatches and comparing with those in the inset, one can deduce,
in each region of the disk, the details of the gravitational lensing.

In Figure \ref{Fig13:DiskPaintSwatch}, the lower disk image wraps under the black hole's shadow and then swings inward, becoming very thin, then up over the shadow and back down and outward to close on itself.  This entire image comes from light rays emitted by the 
disk's bottom face: the wide bottom portion of the image, from rays that originate behind the hole, and
travel under the hole and back upward to the camera; the narrow top portion, from rays
that originate on the disk's front underside and travel under the hole, upward on its back
side, over its top, and down to the camera---making one full loop around
the hole.

There is a third disk image whose bottom portion is barely visible near the shadow's edge.
That third image consists of light emitted from the disk's top face, that travels around the
hole once for the visible bottom part of the image, and one and a half times for the 
unresolved top part of the image.

\begin{figure}[b!]
\begin{center}
\includegraphics[width=0.8\columnwidth]{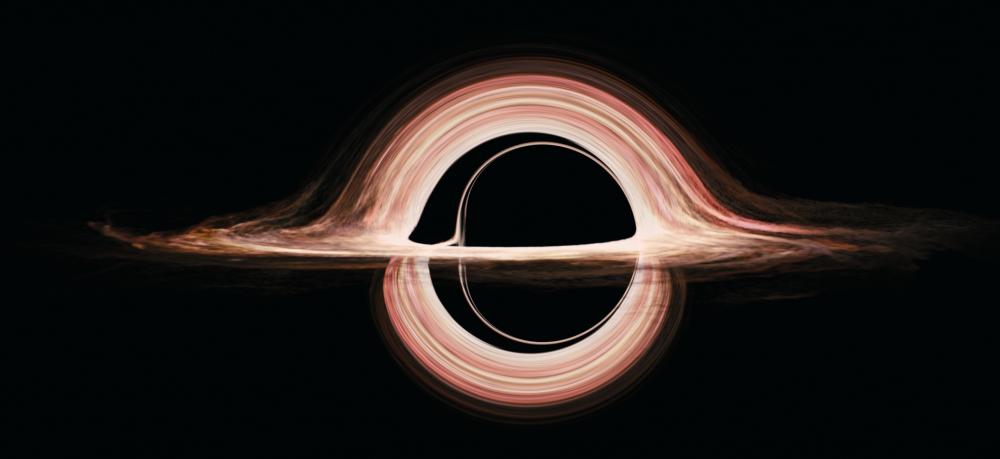}
\end{center}
\caption{A 
moderately realistic accretion disk, created by Double Negative artists and gravitationally
lensed by the same black hole with $a/M=0.999$ as in Figure \ref{Fig13:DiskPaintSwatch} and with the same geometry. 
}
\label{Fig14:DiskNatural999}
\end{figure}

In the remainder of this section \ref{sec:Disk} we  deal with a moderately realistic accretion disk---but a disk created for \emph{Interstellar}  
by Double Negative artists  rather than created by solving 
astrophysical equations such as \cite{NovikovThorne}.  In \ref{subsec:DNGRdisks}
we  give some
details of how this and other Double Negative accretion disk images were created.  This artists' \emph{Interstellar} disk was chosen to be very anemic compared to the disks that astronomers see around black holes and that astrophysicists model --- so the humans who travel near it will not get fried by X-rays and gamma-rays.   It is physically thin and marginally optically thick and lies in the black hole's equatorial plane.  It is not currently accreting onto the black hole, and it has cooled to a position-independent temperature $T=4500$K, at which it emits a black-body spectrum.

\begin{figure}
\begin{center}
\includegraphics[width=0.8\columnwidth]{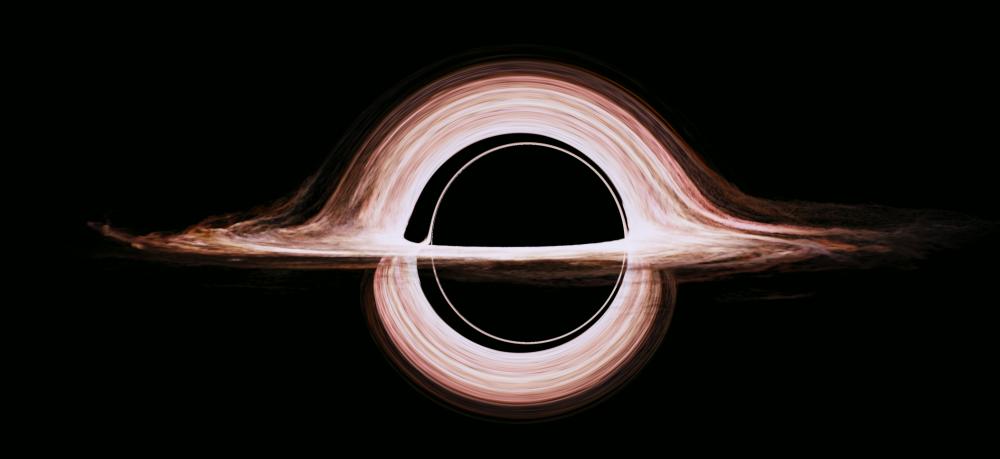}
\vskip1pc
\includegraphics[width=0.8\columnwidth]{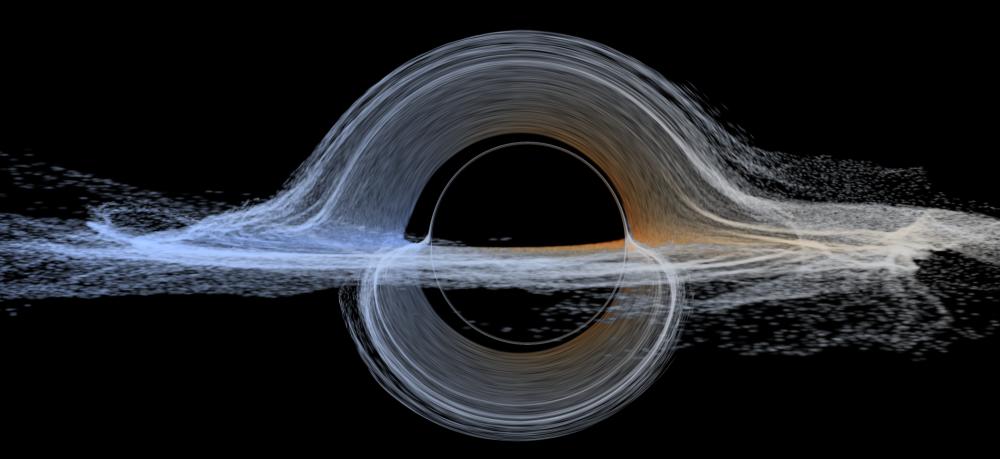}
\vskip1pc
\includegraphics[width=0.8\columnwidth]{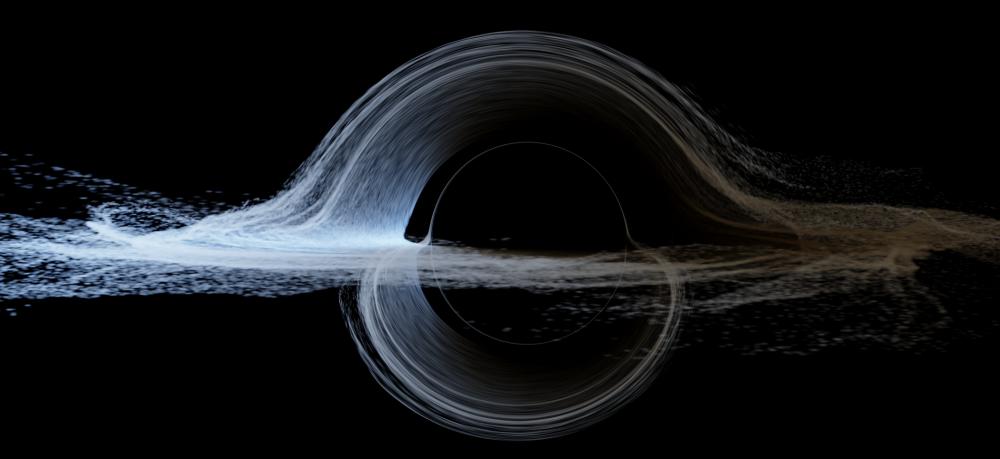}
\end{center}
\caption{(a) The
moderately realistic accretion disk of Figure \ref{Fig14:DiskNatural999} but with the
black hole's spin slowed from $a/M=0.999$ to $a/M=0.6$ for reasons discussed in the text.
(b) This same disk with its colours (light frequencies $\nu$) Doppler shifted and gravitationally
shifted.  (c) The same disk with its specific intensity (brightness) also shifted in accord with
Liouville's theorem, $I_\nu \propto \nu^3$.  This image is what the disk would truly
look like to an observer near the black hole. }
\label{Fig15:Disk}
\end{figure}

 Figure \ref{Fig14:DiskNatural999} shows an image of this artists' disk,  generated with a 
gravitational lensing geometry and computational procedure  identical to those for our paint-swatch disk, Figure \ref{Fig13:DiskPaintSwatch} (no frequency shifts or associated
colour 
and brightness changes; no lens flare).  Christopher Nolan and Paul Franklin decided that the flattened left edge of the black-hole shadow, and the multiple disk images alongside that left edge, and the off-centred disk would be too confusing for a mass audience.  So---although \emph{Interstellar}'s black hole had to
spin very fast to produce the huge time dilations seen in the movie---for visual purposes Nolan and Franklin slowed the spin to $a/M=0.6$, resulting in the disk of  Figure  \ref{Fig15:Disk}a.

 \subsubsection{Colour and brightness changes due to frequency shifts}
 \label{subsec:FrequencyShifts}
 
 The influences of Doppler and gravitational frequency shifts on the appearance of
 this disk are shown in Figures \ref{Fig15:Disk}b,c. 
 
 Since the left side of the disk is moving
 toward the camera and the right side away with speeds of roughly $0.55c$, their light frequencies get shifted blueward on the left and redward on the right---by multiplicative factors of order 1.5 and 0.4 respectively when one combines the Doppler shift with a
 $\sim 20$ percent gravitational redshift.   These frequency changes induce changes in the disk's perceived \emph{colours} (which we compute by convolving the frequency-shifted spectrum with the sensitivity curves of motion picture film) and also induce changes in the disk's perceived \emph{brightness}; see \ref{subsec:DNGRdisks} for some details.  
  
In Figure \ref{Fig15:Disk}b, we have turned on the colour changes,  but not the corresponding
 brightness changes.    As expected, the disk has become blue on the left and red on the right.

In Figure \ref{Fig15:Disk}c, we have turned on both the colour and the brightness changes.  Notice that the disk's left side, moving toward the camera, has become very
 bright, while the right side, moving away, has become very dim. This
 is similar to astrophysically observed jets, emerging from distant galaxies and
 quasars; one jet, moving toward Earth is typically bright, while the other, moving
 away, is often too dim to be seen.
 
 \subsection{Lens flare and the accretion disk in the movie \emph{Interstellar}}
 \label{subsec:DiskLensFlare}
 
 Christopher Nolan, the director and co-writer of {\it Interstellar},
 and Paul Franklin, the visual effects supervisor, were committed to
 make the film as scientifically accurate as possible---within constraints of not confusing
 his mass audience unduly and using images that are exciting and fresh.  A fully realistic accretion disk, Figure  \ref{Fig15:Disk}c,  that is exceedingly
 lopsided, with the hole's shadow barely discernible, was obviously  unacceptable.
 
  \begin{figure}[t!]
\begin{center}
\includegraphics[width=0.9\columnwidth]{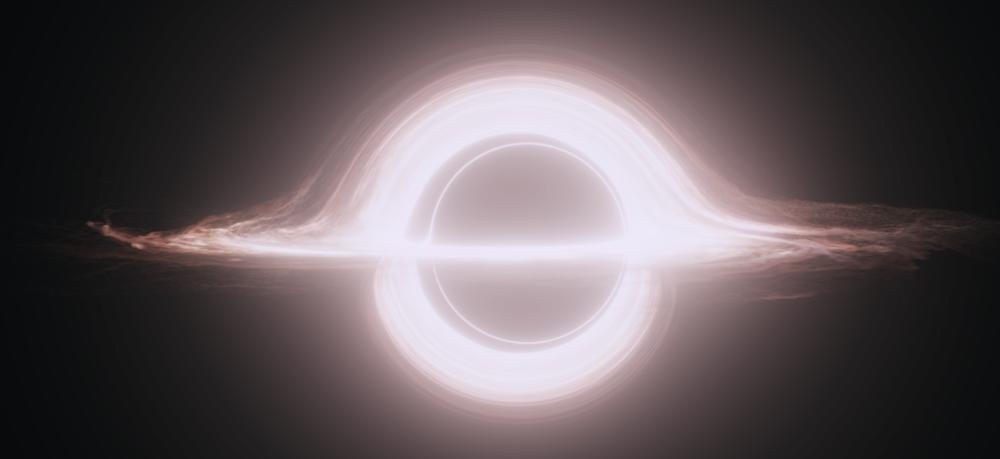}
\end{center}
\caption{The accretion disk of Figure  \ref{Fig15:Disk}a (no colour or brightness
shifts) with lens flare added---a type of lens flare called ``veiling flare'', which has the
look of a soft glow and is very characteristic of IMAX camera lenses. This is
a variant of the accretion disk seen in \emph{Interstellar}.  (Figure created  by our Double Negative team using DNGR,  and 
TM \& \copyright Warner Bros. Entertainment Inc. (s15))
 This image may be used under the terms
of the Creative Commons Attribution-NonCommercial-NoDerivs 3.0 (CC BY-NC-ND
3.0) license. Any further distribution of these images must maintain attribution to the
author(s) and the title of the work, journal citation and DOI. You may not use the
images for commercial purposes and if you remix, transform or build upon the images,
you may not distribute the modified images.
}
\label{Fig16:DiskLensFlare}
\end{figure}
 
The first image in Figure \ref{Fig15:Disk}, the one
 without frequency shifts and associated colour and brightness changes,  was particularly appealing, 
but it lacked one element of realism that few  astrophysicists would ever think
 of (though astronomers take it into account when modelling their
 own optical instruments).  Movie audiences are accustomed to seeing scenes filmed through a real camera---a
 camera whose optics scatter and diffract the incoming light, producing what is called
 \emph{lens flare}.  As is conventional for movies (so that computer generated
 images will have visual continuity with images shot by real cameras), Nolan and Franklin asked that simulated lens flare be imposed on the 
 accretion-disk image.  The result, for the first image in Figure \ref{Fig15:Disk},
 is Figure \ref{Fig16:DiskLensFlare}.

 This, with some embellishments, is the accretion disk seen around the black hole
 Gargantua in \emph{Interstellar}.  

\emph{All} of the black-hole and accretion-disk images in \emph{Interstellar} were generated using DNGR, with a single exception:  When Cooper (Matthew McConaughey), riding in the Ranger spacecraft, has plunged into the black hole Gargantua, the camera, looking back upward from inside the event horizon, sees the gravitationally distorted external universe within the accretion disk and the black-hole shadow outside it --- as general relativity predicts.  Because DNGR uses Boyer-Lindquist coordinates, which do not extend smoothly through the horizon, this exceptional image had to be constructed by Double Negative artists manipulating DNGR images by hand.

  \subsection{Some details of the DNGR accretion-disk simulations}
 \label{subsec:DNGRdisks}

 \subsubsection{Simulating lens flare}
 \label{subsec:DNGRFlare}
 
In 2002 one of our authors (James)  formulated and perfected the following (rather obvious) method for applying
lens flare to images.
The appearance of a distant star on a camera's focal plane is mainly determined by the point spread function of the camera's optics. For Christopher Nolan's films we measure the point spread function by recording with HDR photography (see e.g. \cite{REINHARD05}) a point source of light with the full set of 35mm and 65mm lenses
 typically used in his IMAX and anamorphic cameras, attached to a single lens reflex camera.
We apply the camera's lens flare to an image by
 convolving it with this point spread function.  (For these optics concepts
 see, e.g., \cite{BROOKERG}.) For the image \ref{Fig15:Disk}a, this
 produces Figure \ref{Fig16:DiskLensFlare}. More recent work \cite{HULLIN} does a more thorough analysis of reflections
between the optical elements in a lens, but requires detailed knowledge of each lens' construction, which was not
readily available 
for our \emph{Interstellar} work.
 
 \subsubsection{Modelling the accretion disk for \emph{Interstellar}}
 \label{subsec:DNGRInterstellarDisk}
 
 As discussed above, the accretion disk in \emph{Interstellar} was an artist's conception,
 informed by images that astrophysicists have produced, rather than computed directly
 from astrophysicists' accretion-disk equations such as \cite{NovikovThorne}.
 
 In our work on \emph{Interstellar}, we developed three different types of disk models:
\begin{itemize}
\item
an infinitely thin, planar disk, with colour and optical thickness defined by an artist's image;
\item a three dimensional `voxel' model;
\item
for close-up shots, a disk with detailed texture added by modifying a
commercial renderer, Mantra \cite{MANTRA}.
\end{itemize}
We discuss these briefly in \ref{subsec:DNGRdisks}.

\section{Conclusion}
\label{sec:Conclusion}

In this paper we have described the code DNGR, developed at Double Negative
Ltd., for creating general relativistically correct images of black holes
and their accretion disks.  We have described our use of DNGR to generate 
the accretion-disk images
seen in the movie \emph{Interstellar}, and to explain effects that influence  the disk's
appearance: light-ray bending, Doppler and gravitational frequency shifts, 
shift-induced colour and brightness changes, and camera lens flare.

We have also used DNGR to explore, for a camera orbiting a fast spinning black hole,  the gravitational lensing of a star field on the celestial sphere---including the lensing's
caustics and critical curves,  and how they influence the stellar images' 
pattern on the camera's sky, the creation and
annihilation of image pairs, and the image motions.  

Elsewhere \cite{DNGRwormhole} we describe our use of DNGR to explore gravitational lensing by hypothetical wormholes; particularly, the influence of a wormhole's 
length and shape on its lensing of stars and nebulae; and we describe the choices of 
length and shape  that were made for \emph{Interstellar}'s wormhole and how we 
generated
that movie's wormhole images.

\ack

For very helpful advice during the development of DNGR and/or during the research with it
reported here, we thank  Christopher Nolan, Alain Riazuelo, Jean-Pierre Luminet, Roger Blandford, Andy Bohn, Francois Hebert, William Throwe, Avery Broderick and Dimitrios Psaltis.  
For contributions to DNGR and its applications, we thank  members of the Double Negative R\&D team
Sylvan Dieckmann, Simon Pabst, Shane Christopher, Paul-George Roberts, and Damien Maupu; and also Double Negative artists
Fabio Zangla, Shaun Roth, Zoe Lord, Iacopo di Luigi, Finella Fan, Nicholas New, Tristan Myles, and Peter Howlett.

The construction of   DNGR was funded by Warner Bros.\ Entertainment Inc., for generating visual effects for the movie \emph{Interstellar}.  We thank Warner Bros.\ for authorising this code's additional use for scientific research, and in particular the
research reported in this paper.

\appendix

\section{Some Details of DNGR}
\label{sec:App}

\subsection{Ray-tracing equations}
\label{subsec:App-RayTrace}

In this section we give a step-by-step prescription for computing (i)  the ray-tracing map
from a point $\{\theta_{cs}, \phi_{cs}\}$ on the camera's local sky to a point $\{\theta',
\phi'\}$ on the celestial sphere, and also (ii) the  blue shift $f_c/f\,'$ of light emitted from the celestial sphere with frequency $f\,'$ and received at the camera with frequency $f_c$.
Throughout we set the black hole's
mass $M$ to unity, so distances are measured in terms of $M$.

The foundations for our prescription are: (i) the Kerr metric written in Boyer-Lindquist coordinates\footnote{Our choice of Boyer-Lindquist coordinates prevents the camera from descending into the black hole.  For that, it is not hard to switch to ingoing Kerr coordinates; but if one wants to descend even further, across an inner (Cauchy) horizon of the maximally extended Kerr spacetime, one must then switch again, e.g., to outgoing Kerr coordinates.  Alain Riazuelo (private communication) has created an implementation of the ray tracing equations that does such multiple switches.  As a foundation for \emph{Interstellar}, Thorne did so but only for the switch to ingoing Kerr
coordinates, as  \emph{Interstellar} assumes the inner horizons have been replaced by the Marolf-Ori shock singularity \cite{Marolf-Ori}
and the Poisson-Israel mass-inflation singularity \cite{Poisson-Israel}; see Chapters
24 and 26 of \cite{TSI}.}

\begin{equation}
ds^2 = -\alpha^2 dt^2 + (\rho^2/\Delta)dr^2 + \rho^2 d\theta^2 + \varpi^2 (d\phi - \omega dt)^2
\;,
\end{equation} 
where
\begin{eqnarray}
\rho = \sqrt{r^2 + a^2 \cos^2\theta}\;, \;\;
\Delta = r^2 -2r+a^2\;, \;\;
\Sigma=\sqrt{(r^2+a^2)^2 -a^2 \Delta \sin^2\theta}\;, \nonumber \\
\alpha={\rho \sqrt{\Delta}\over\Sigma}\;, \quad
\omega = {2 a r \over \Sigma^2} \;, \quad
\varpi =  {\Sigma \sin\theta \over \rho}\;;
\label{eq:KerrQuantities}
\end{eqnarray}
(ii) the 3+1 split of spacetime into space plus time embodied in this form of the metric; and (iii) 
the family of fiducial observers (FIDOs) whose world lines are orthogonal to the
3-spaces of constant $t$, and their orthonormal basis vectors that lie in those 3-spaces
and are depicted in Figure \ref{Fig1:SkyMapA},
\begin{equation}
{\bf e}_{\hat r} = {\sqrt{\Delta}\over\rho} {\partial\over\partial r}\;, \quad
{\bf e}_{\hat \theta} = {1\over\rho}{\partial\over\partial \theta}\;, \quad
{\bf e}_{\hat \phi} = {1\over \varpi}{\partial\over\partial \phi}\;.
\label{eq:FIDObasis}
\end{equation}
We shall also need
three functions of $r$ and of a ray's constants of motion $b$ (axial angular momentum)
and $q$ (Carter constant), which appear in the ray's evolution equations (\ref{eq:Rays2}) below---i.e., in the equations for a \emph{null} geodesic:
\begin{equation}
\fl
P = r^2+a^2 - a b\;,\;\;
R = P^2 - \Delta[(b-a)^2+q]\;, \;\;
\Theta = q-\cos^2\theta\left({b^2\over\sin^2\theta} - a^2\right)\;.
\label{eqs:PRTheta}
\end{equation}
And we shall need the function $q_o(b_o)$, which identifies the constants of motion for rays (photons) that are unstably trapped in constant-$r$ orbits around the black hole.  This function
is defined parametrically by \cite{Teo}
\begin{equation}
b_o  \equiv -{r_o^3-3 r_o^2 + a^2 r_o + a^2 \over a(r_o-1)} \;,\;\;
q_o \equiv - {r_o^3(r_o^3 - 6 r_o^2 + 9 r_o - 4a^2)\over a^2(r_o-1)^2}\;,
\label{eq:boqo}
\end{equation}
where $r_o$ is the radius of the trapped orbit, which runs over the interval 
$r_1  \le  r_o \le r_2 $, with
\begin{equation}
r_1 \equiv 2 \left\{1 + \cos\left[{2\over3} \arccos(-a)\right]\right\}\;, \;\;
r_2  \equiv 2 \left\{1 + \cos\left[{2\over3} \arccos(+a)\right]\right\}\;. 
\label{eq:r1r2}
\end{equation}

The prescription that we use, in DNGR, for computing the ray-tracing map and the
blue shift is the following concrete embodiment of the discussion in Section \ref{subsec:RayTracing}.

\begin{enumerate}

\item
Specify the camera's location $(r_c,\theta_c,\phi_c)$, and its speed $\beta$ and the
components $B_{\hat r}$, $B_{\hat \theta}$, and $B_{\hat \phi}$ of its direction of motion relative to
the FIDO at its location; and specify the ray's incoming direction $(\theta_{cs}, \phi_{cs})$ on the camera's local sky.  [Note: If the camera is in a circular, equatorial geodesic orbit around the black hole, then 
\begin{equation}
\fl
B_{\hat r} = 0\;,\quad B_{\hat \theta}=0\;, \quad B_{\hat \phi}=1\;, \quad
\beta = {\varpi\over\alpha}(\Omega - \omega)\;, \quad \textrm{where } \;\Omega = {1\over(a+r_c^{3/2})}
\label{eq:CircGeod}
\end{equation} 
is the 
geodesic angular velocity at the camera's radius $r_c$, and the other quantities are defined in Equations (\ref{eq:KerrQuantities}).]

\item
Compute, in the camera's proper reference frame, the Cartesian components (Figure \ref{Fig1:SkyMapA}) of the unit vector $\bf N$ that points in the direction of the incoming ray
\begin{equation}
N_x = \sin\theta_{cs} \cos\phi_{cs}\;, \;
N_y = \sin\theta_{cs} \sin\phi_{cs}\;, \;
N_z = \cos\theta_{cs}\;.
\label{eq:CameraAngles}
\end{equation}

\item
Using the equations for  relativistic aberration, compute the direction of motion
of the incoming ray, ${\bf n}_F$, as measured by the FIDO in Cartesian coordinates aligned with those of
the camera:
\begin{equation}
n_{F\,y}  = {-N_y + \beta \over 1-\beta N_y}\;, \;\;
n_{F\,x} = {- \sqrt{1-\beta^2} N_x \over 1-\beta N_y}\;,\;\;
n_{F\,z} = {- \sqrt{1-\beta^2} N_z \over 1-\beta N_y}\;;
\label{eq:aberrationA}
\end{equation}
and from these, compute the components of ${\bf n}_F$ on the FIDO's spherical
orthonormal basis:
\begin{eqnarray}
\fl 
n_{F\,\hat r} ={B_{\hat \phi}\over\kappa} \,n_{F\,x} + B_{\hat r} \,n_{F\,y} + {B_{\hat r} B_{\hat \theta}\over\kappa}
\,n_{F\,z}\;, \quad
n_{F\,\hat \theta} = B_{\hat \theta} \,n_{F\,y} - \kappa \,n_{F\,z}\;, \nonumber \\
\fl 
n_{F\, \hat \phi} = -{B_{\hat r}\over\kappa}\, n_{F\, x} + B_{\hat \phi} \, n_{F\,y} + {B_{\hat \theta} B_{\hat \phi}
\over \kappa}\, n_{F\, z}\;, \quad
\textrm{where } \kappa \equiv \sqrt{1-B_{\hat \theta}^2} = \sqrt{B_{\hat r}^2+B_{\hat \phi}^2}\;.
\label{eq:kappa}
\end{eqnarray}
\label{eq:nFtransform}

\item
Compute the ray's canonical momenta (covariant coordinate components of 
its 4-momentum) with its conserved energy $-p_t$ set to unity as a convenient convention:
\begin{eqnarray}
p_t =-1\;, \quad
p_r = E_F {\rho \over\sqrt{\Delta}} \,n_{F\,\hat r}\;, \quad
p_\theta = E_F \rho \,n_{F\,\hat\theta}\;, \quad
p_\phi = E_F \varpi \, n_{F\,\hat\phi}\;, \quad \nonumber \\
\textrm{where } E_F = {1\over \alpha+\omega\varpi n_{F\,\hat\phi}}
\label{eq:canonicalmomenta}
\end{eqnarray}
is the energy measured by the FIDO.
(Note: $p_\alpha$ can also be regarded as the ray's wave vector or simply
as a tangent vector to the ray.)  Then compute the
ray's other two conserved quantities: its axial angular momentum $b$ and its
Carter constant $q$:
\begin{equation}
b =  p_\phi\;, \quad
q = p_\theta^2 + \cos^2\theta\left({b^2\over\sin^2\theta} - a^2 \right) \;.
\label{eq:bq}
\end{equation}
(Note: if we had not set $p_t=-1$, then $b$ would be $-p_\phi/p_t$ and 
$q$ would be (Carter constant)$/p_t^2$.)

\item
Determine, from the ray's constants $\{b,q\}$, whether it comes from the horizon or the celestial sphere by the following
algorithm: (a) Are both of the following conditions satisfied?: (v.1)
\begin{equation}
b_1 < b < b_2\;, \;\;  \textrm{where } b_1=b_o(r_2)\;, \; \;
b_2 = b_o(r_1)\;, 
\label{eq:b1b2}
\end{equation}
with $r_1$ and $r_2$ given by Equations  (\ref{eq:r1r2}) and the function
$b_o(r_o)$  given by Equations (\ref{eq:boqo}); and (v.2)  the ray's value of $q$
lies in the range
\begin{equation}
q< q_o(b)\;, \;\; \textrm{where } q_o(b) \textrm{ is defined  by Equations (\ref{eq:boqo}).}
\label{eq:qob}
\end{equation}
(b) If the answer to (a) is \emph{yes}, then there are no radial turning points for that $\{b,q\}$, whence  if $p_r>0$ at
the camera's location, the ray comes from the horizon; and if $p_r<0$ there,
it comes from the celestial sphere.  (c) If the answer to (a) is \emph{no}, then there
are two radial turning points for that $\{b,q\}$, and if the camera radius $r_c$ is greater 
than or equal to the radius $r_{\rm up}$ of the upper turning point, then the ray comes from
the celestial sphere; otherwise, it comes from the horizon.  Here
$r_{\rm up}$ is the largest real root
of $R(r)=0$, with $R(r)$  defined by Equations (\ref{eqs:PRTheta}).
\item
 If the ray comes from the celestial sphere, then  compute its point of origin there as follows: 
Beginning with the computed values of the constants of motion and canonical momenta,
and beginning in space at the camera's location, 
integrate the ray equations\footnote{These are the super-Hamiltonian variant of the
null geodesic equations; see, e.g., Section 33.5 of \cite{MTW} or Appendix A of 
\cite{Levin}.}
\begin{eqnarray}
{dr\over d\zeta} = {\Delta\over \rho^2}\, p_r\;, \quad
{d\theta\over d\zeta} = {1\over\rho^2}\, p_\theta\;, \quad
{d\phi\over d\zeta} = 
- {\partial \over \partial b}\left({R+\Delta\Theta\over 2 \Delta \rho^2}\right)\;, \nonumber \\
{d p_r\over d\zeta} =  {\partial \over \partial r}
\left[- {\Delta\over 2\rho^2}\, p_r^2 
- {1\over 2\rho^2}\, p_\theta^2 + \left({R+\Delta\Theta\over 2 \Delta \rho^2}\right) \right] \;, \nonumber \\
{d p_\theta\over d\zeta} =  {\partial \over \partial \theta}\left[- {\Delta\over 2\rho^2}\, p_r^2 
- {1\over 2\rho^2} \,p_\theta^2 + \left({R+\Delta\Theta\over 2 \Delta \rho^2}\right) \right] \;
\label{eq:Rays2}
\end{eqnarray}

numerically, backward in time 
from $\zeta= 0$ to $\zeta = \zeta_f$, where $\zeta_f$ is either $-\infty$ or some very negative value.  (This super-Hamiltonian version of the ray equations is 
well behaved and robust at turning points, by contrast, for example, with a
commonly used version of the form $\rho^2 dr/d\zeta = \pm \sqrt{R}$, 
$\rho^2 d\theta/d\zeta=\pm \sqrt{\Theta}$, ... ; Eqs.\ (33.32) of MTW 
\cite{MTW}.) 
The ray's source point 
$(\theta',\phi')$ on the
celestial sphere is $\theta'=\theta(\zeta_f)$, $\phi'=\phi(\zeta_f)$.
\item
Compute the light's blue shift, i.e.\ the ratio of its frequency $f_c$ as seen by the camera to
its frequency $f'$ as seen by the source at rest on the celestial sphere, from  \begin{equation}
{f_c\over f'} 
=
 \left(\frac{\sqrt{1-\beta^2}}{1- \beta N_y }\right)
 \left( {1-b\omega\over\alpha}\right) \;.
 \label{eq:FinalBlueShift}
 \end{equation}
\end{enumerate}

\subsection{Ray-bundle equations}
\label{subsec:App-RayBundle}

In this section we give a step-by-step prescription for evolving a tiny bundle of rays
(a light beam) along a reference ray, from the camera to the celestial sphere, with
the goal of learning the beam's major and minor angular diameters $\delta_+$
and $\delta_-$ on the celestial sphere, and the angle $\mu$ from ${\bf e}_{\theta'}$ on
the celestial sphere to the beam's major axis (cf.\ Figure\ \ref{Fig1:SkyMapA}).  As in the previous subsection, we
set the black hole's mass to unity, $M=1$.

Our prescription is a concrete embodiment of the discussion in Section \ref{subsec:RayBundle}; it is a variant of the Sachs optical scalar equations
\cite{Ehlers};  and it is a slight modification of a prescription developed
in the 1970s by Serge Pineault and Robert Roeder 
\cite{PR1,PR2}.\footnote{The Pineault-Roeder equations
(which were optimised for rays that begin near the black hole and travel from there to earth) become asymptotically singular for our situation, where the rays begin on the celestial sphere and travel to a camera near the black hole.     We have reworked the Pineault-Roeder analysis so as to make their equations fully nonsingular and robust for our situation. This reworking entailed
changing their null vector $l^{(a)}$ from $2^{-1/2}(1,-1,0,0)$ to $2^{-1/2}(1,+1,0,0)$ in
Equation (6) of \cite{PR1}.  This simple change of one foundational equation produced many changes in the final evolution equations for the ray bundle.  Most 
importantly, it changed $1/(p^{\hat t}+p^{\hat r})$ to $1/(p^{\hat t}-p^{\hat r})$ in Eqs.\ (\ref{eq:RePsios}) and (\ref{eq:ImPsios}); 
the former diverges at the celestial sphere for our ingoing ray bundles, while
the latter is finite there.  For a detailed derivation of our prescription, see
\cite{KipMathematica}.}

Our prescription relies on the following functions, which are defined
along the reference ray.  (a) The components of the ray bundle's 4-momentum (wave vector) on
the FIDO's orthonormal spherical basis
\begin{equation}
\fl
p^{\hat r} = p_{\hat r} ={\sqrt{\Delta}\over \rho} p_r\;,\quad
p^{\hat \theta} = p_{\hat \theta} = {p_\theta \over\rho} \;,\quad
p^{\hat \phi} = p_{\hat \phi} = {p_\phi \over\varpi} = {b\over\varpi}\;,\quad
p^{\hat t} = \sqrt{p_{\hat r}^2 +p_{\hat \theta}^2 + p_{\hat \phi}^2}\;;
\label{eq:kcomponents}
\end{equation}
(b) A function $\mathcal M$ defined on the phase space of the reference ray by

 \begin{eqnarray}
\fl
\mathcal M =
{1\over {4 \rho ^3 \Sigma ^2
   (p^{\hat r}-p^{\hat t})}}
\Bigg\{8 a^3 \sqrt{\Delta } r \sin ^2\theta \cos \theta
   \left(p_{\hat \theta}^2-p_{\hat \phi}^2\right) \nonumber\\
\fl \quad
  +2
   (p^{\hat r}-p^{\hat t}) \left\lgroup a \sin \theta \left[a^2
   \left(p_{\hat \theta} \,(a-r) (a+r) \cos 2\theta+2
   \sqrt{\Delta } \,p^{\hat t}\, r \sin 2 \theta
   \right)
   \right.\right. \nonumber\\
\fl \quad\quad
 \left.\left.
 +p_{\hat \theta} \,\left(a^4-3 a^2 r^2-6
   r^4\right)\right]
   -2 p_{\hat \phi}\, \left(a^2+r^2\right)^3
   \cot \theta\right\rgroup
   \label{eq:calM}\\
\fl \quad
   -a^2 p_{\hat \phi}\, \sin \theta
   \left\lgroup 2 \sqrt{\Delta } \,p_{\hat \theta}\, \sin \theta
   \left[a^2 (r-1) \cos 2\theta+a^2 (r-1)+2 r^2
   (r+3)\right]
   \right. \nonumber \\
\fl \quad\quad
    \left.
   +a^2 \Delta  p^{\hat t}\, \cos 3\theta+\cos
   \theta \left[a^4 (7 p^{\hat t}-4 p^{\hat r})+a^2 r
   (p^{\hat t} \,(15 r-22)-8 p^{\hat r} \,(r-2))
\right.\right. \nonumber \\
\fl \quad\quad\quad
   \left.\left.
-4 r^3 (p^{\hat r}\,
   (r-4)-2 p^{\hat t}\, (r-3))\right]\right\rgroup
   \Bigg\} \;. \nonumber 
    \end{eqnarray}
(c) A complex component $\Psi_{0*}$ of the Weyl tensor, evaluated on vectors constructed from the ray's 4-momentum;  its real and imaginary parts work out to be:

\begin{eqnarray}
\fl
\Re(\Psi_{0*}) = {1\over 2 (w-1) (p^{\hat r}-p^{\hat t})^2}
\Bigg\{
Q_1 \left\lgroup-3 w \left(p_{\hat \theta}^2-p_{\hat \phi}^2\right)^2-3 p_{\hat r}^4 w
+2 p_{\hat r}^3 [3 p^{\hat t} \,w+p_{\hat \phi} \,S (w-1)]
\right.\nonumber\\
\fl 
 \left.
    +6 p_{\hat r}^2 \left[p_{\hat \theta}^2-(p^{\hat t})^2\, w-p^{\hat t}\, p_{\hat \phi} \,S (w-1)-p_{\hat \phi}^2 (w+1)\right]
    +2 p^{\hat r} \left[p_{\hat \phi}\, S (w-1) \left(p_{\hat \phi}^2-3 p_{\hat \theta}^2\right)
\right.\right.\nonumber\\   
\fl \left. \left. 
+3 (p^{\hat t})^3 w
   +3 (p^{\hat t})^2 p_{\hat \phi}\, S (w-1)-3 p^{\hat t}\, (w+2)
   \left(p_{\hat \theta}^2-p_{\hat \phi}^2\right)\right]-3 (p^{\hat t})^4 w-2 (p^{\hat t})^3
   p_{\hat \phi}\, S (w-1)
\right. \nonumber\\
\fl \left.
+6 (p^{\hat t})^2 \left[p_{\hat \theta}^2 (w+1)-
   p_{\hat \phi}^2\right]
-2 p^{\hat t}\, p_{\hat \phi} \,S (w-1) \left(p_{\hat \phi}^2-3 p_{\hat \theta}^2\right)\right\rgroup \; + \; 
   2\,  Q_2\,p_{\hat \theta}\, \left\lgroup3 p_{\hat \phi}\, w \left(p_{\hat \phi}^2-p_{\hat \theta}^2\right)\right.
   \nonumber\\
\fl \left.
+p_{\hat r}^3 S(1- w)+3 p_{\hat r}^2 [p^{\hat t} \,S
   (w-1)+p_{\hat \phi}\, (w+2)]   -p_{\hat r}\, \left[-S (w-1) \left(p_{\hat \theta}^2-3
   p_{\hat \phi}^2\right) +3 (p^{\hat t})^2 S (w-1)
\right. \right.\nonumber\\
\fl \left.\left.
+6 p^{\hat t} \,p_{\hat \phi}\,
   (w+2)\right]
+(p^{\hat t})^3 S (w-1)+3 (p^{\hat t})^2 p_{\hat \phi} \,(w+2)-p^{\hat t}\, S (w-1)
   \left(p_{\hat \theta}^2-3 p_{\hat \phi}^2\right)\right\rgroup
   \Bigg\}  \;,
\label{eq:RePsios}
    \end{eqnarray}

\begin{eqnarray}
\fl
\Im(\Psi_{0*}) = {1\over (w-1) (p^{\hat r}-p^{\hat t})^2}
\Bigg\{
-Q_2 \left\lgroup 6 p_{\hat \theta}^2 p_{\hat \phi}^2 w+p_{\hat r}^3 [3 p^{\hat t}\,
   w+p_{\hat \phi} \,S (w-1)]
   \right. \nonumber\\
\fl \left.
+3 p_{\hat r}^2 \left[p_{\hat \theta}^2 (w+1)-2 (p^{\hat t})^2 w     
   -p^{\hat t}\, p_{\hat \phi} \,S (w-1)-p_{\hat \phi}^2\right]+p_{\hat r}\, \left[p_{\hat \phi} \,S (w-1) \left(p_{\hat \phi}^2-3 p_{\hat \theta}^2\right) \right. \right. \nonumber\\
\fl \left.\left. 
+3 (p^{\hat t})^3 w+3
   (p^{\hat t})^2 p_{\hat \phi}\, S (w-1)
   -3 p^{\hat t} \,(w+2) \left(p_{\hat \theta}^2-p_{\hat \phi}^2\right)\right]-(p^{\hat t})^3 p_{\hat \phi} \,S (w-1)\right\rgroup
\nonumber\\
\fl
+3 (p^{\hat t})^2
   \left[p_{\hat \theta}^2-p_{\hat \phi}^2 (w+1)\right]-p^{\hat t}\, p_{\hat \phi} \,S
   (w-1) \left(p_{\hat \phi}^2-3 p_{\hat \theta}^2\right)
   +Q_1\,p_{\hat \theta}\,  \left\lgroup3 p_{\hat \phi} \,w \left(p_{\hat \phi}^2-p_{\hat \theta}^2\right)
   \right. \\
\fl
   +p_{\hat r}^3 S(1- w)+3 p_{\hat r}^2 [p^{\hat t} \,S (w-1)+p_{\hat \phi}\,
   (w+2)]-p_{\hat r}\, \left[-S (w-1) \left(p_{\hat \theta}^2-3 p_{\hat \phi}^2\right)
      +3 (p^{\hat t})^2 S (w-1)
   \right. \nonumber \\
\fl
\left. \left. 
+6 p^{\hat t} \,p_{\hat \phi} \,(w+2)\right]+(p^{\hat t})^3 S (w-1)+3
   (p^{\hat t})^2 p_{\hat \phi} \,(w+2)-p^{\hat t} \,S (w-1) \left(p_{\hat \theta}^2-3
   p_{\hat \phi}^2\right)\right\rgroup   \Bigg\}  \;. \nonumber
\label{eq:ImPsios}
    \end{eqnarray}
Here $Q_1$, $Q_2$, $w$, and $S$ are given by
\begin{eqnarray}
Q_1 = {r(r^2 - 3 a^2 \cos^2\theta)\over \rho^6}\;,\quad
Q_2 ={ a \cos\theta(3 r^2-a^2 \cos^2\theta) \over \rho^6}\;,\nonumber \\
w= {\Delta a^2 \sin^2\theta\over (r^2+a^2)^2}\;, \quad
S = {3 a \sqrt{\Delta}(r^2+a^2)\sin\theta \over \Sigma^2}\;.
\label{eq:Psios}
\end{eqnarray}

We shall state our prescription for computing the shape and orientation of the
ray bundle on the celestial sphere separately, for two cases: a ray bundle that
begins circular at the camera; and one that begins elliptical.  Then,  we shall
briefly sketch the Pineault-Roeder \cite{PR1,PR2} foundations for these prescriptions.

\subsubsection{Circular ray bundle at camera}
\label{subsec:CircAtCamera}

If the ray bundle begins at the camera with a tiny circular angular diameter
$\delta_{cs}$ as measured by the camera, then our prescription for computing
its shape and orientation at the celestial sphere is this:
\begin{enumerate}
\item
Introduce five new  functions along the reference ray: 
$u(\zeta)$, $v(\zeta)$, $g(\zeta)$, $h(\zeta)$, and $\chi(\zeta)$ 
[which are denoted $x$, $y$, $p$, $q$ and $\chi$ by Pineault and Roeder].  Give them and their
derivatives the following initial conditions at the camera, $\zeta=0$:
\begin{equation}
u, \, v, \, g, \, h, \, \dot v, \, \dot g, \, \dot h, \, \dot \chi \;\; \textrm{all vanish}\;; \quad
 \dot u = 1\;.
\end{equation}
Here a dot means $d/d\zeta$.
\item
Integrate the following differential equations backward along the ray toward its source
at the celestial sphere, i.e. from $\zeta=0$ to $\zeta_f$.  These are coupled
to the ray equations (\ref{eq:Rays2}) for the reference ray.
\begin{eqnarray}
\fl
\ddot u = - \Psi (g \cos\psi + h \sin\psi)\;, \quad
\ddot v = - \Psi(g\sin\psi-h \cos\psi)\;,\nonumber \\
\fl
\ddot g = - \Psi(u\cos\psi + v \sin\psi)\;, \quad
\ddot h = - \Psi(u\sin\psi - v\cos\psi)\;, \quad
\dot \chi = \mathcal M\;.
\label{eq:BundleEvolution}
\end{eqnarray}
Here 
\begin{equation}
\fl
\Psi = | \Psi_{0*} | = \sqrt{[\Re(\Psi_{0*})]^2 + [\Im(\Psi_{0*})]^2}\;, \quad
\psi = \arg(\Psi_{0*} ) -2\chi \;.
\end{equation}
\item
Upon reaching the celestial sphere at $\zeta = \zeta_f$, evaluate the angular diameters of the
bundle's major and minor axes using the equations
\begin{equation}
\fl
\delta_+ = \delta' (\sqrt{\dot u^2+\dot v^2} + \sqrt{\dot g^2+\dot h^2})\;, \quad
\delta_- = \delta' (\sqrt{\dot u^2 +\dot v^2} - \sqrt{\dot g^2+\dot h^2})\;. 
\label{eq:deltapm}
\end{equation}
Here 
$\delta'$ is the angular radius that the bundle would have 
had at the celestial sphere if
spacetime curvature had not affected it:
\begin{equation}
 \delta' = \delta_{cs}
 \left({\sqrt{1-\beta^2} \over 1-\beta N_y } \right)
\left({1-b\omega \over \alpha}\right)_c\;,
\label{eq:deltaprime}
\end{equation}
where the right hand side is to be evaluated at the camera.
Also evaluate the bundle's orientation angle at the celestial sphere from
\begin{equation}
\mu = \chi +{1\over2}\arg[(\dot u+i\dot v)(\dot g + i \dot h)]\;,
\label{eq:mu}
\end{equation}
with the right hand sides evaluated at the celestial sphere, $\zeta=\zeta_f$,
and with $i = \sqrt{-1}$.  The first term, $\chi$, is the contribution from parallel
transporting the bundle's major axis along the reference ray; the second term
is the deviation from parallel transport caused by spacetime curvature.

\end{enumerate}

\subsubsection{Elliptical ray bundle at camera}

Our prescription, for a ray bundle that begins elliptical at the camera, is this:

\begin{enumerate}
\item
In the camera's proper reference frame, specify 
the Cartesian components of the bundle's major axis
$\{J_x,J_y,J_z\}$ (a unit vector), subject to the constraint that
$\bf J$  be orthogonal to the ray direction ${\bf n} = - {\bf N}$  [Equation (\ref{eq:CameraAngles})], 
$J_x n_x +  J_y n_y + J_z n_z = 0$.
\item
Compute the major-axis direction in the FIDO frame using the
transformation equations 
\begin{equation}
\fl
J_{F\,y}  = {\sqrt{1-\beta^2} J_y  \over 1+\beta n_y}\; \quad
J_{F\,x} = J_x - {\beta n_x J_y \over 1+\beta n_y}\; \quad
J_{F\,z} =J_z -{\beta n_z  J_y\over 1+\beta n_y}\;,
\label{eq:MajorAxisTransform}
\end{equation}
where $\beta$ is the camera's speed (in the $y$ direction) relative to the FIDO
at its location.
\item
In the camera's reference frame, specify (a) 
the bundle's angular diameter $\delta_{cs}$ \emph{along its major axis}, and (b) 
the ratio $e$ of the angular diameter along the
minor axis to that along the major axis.  These quantities are the same in the FIDO frame
as in the camera frame: they are insensitive to special relativistic aberration.
\item
Then the initial conditions for the integration variables $u$, $v$, $g$, $h$,
$\chi$ are
\begin{eqnarray}
\fl
 u,v,g,h,\chi,\dot\chi \; \;\textrm{all vanish;} \quad
 \dot u = \mb a \cdot \mb J_F {(1+e)}/2\;,  \quad
\dot v = \mb b \cdot \mb J_F {(1+e)}/2\;, \nonumber \\
\dot g = \mb a \cdot \mb J_F {(1-e)}/2\;, \quad
\dot h = \mb b \cdot \mb J_F {(1-e)}/2\;. 
\label{eq:InitialConditions}
\end{eqnarray}
Here  $\mb a \cdot \mb J_F = a_x J_{F\,x} + a_y J_{F\,y}
+ a_z J_{F\,z}$ is the Cartesian inner product, and $\mb a$ and $\mb b$ are  
two orthogonal unit vectors that span the plane orthogonal to the fiducial ray:
\begin{eqnarray}
a_{\hat r} = n_{F\,\hat\theta}  \;, \quad
a_{\hat \theta}=1 -{n_{F\,\hat\theta} ^2\over(1-n_{F\,\hat r} )}\;,\quad
a_{\hat \phi} = - {n_{F\,\hat\theta} n_{F\,\hat\phi}  \over (1-n_{F\,\hat r})}\;; \nonumber
\end{eqnarray}
\begin{eqnarray}
b_{\hat r}=n_{F\,\hat\phi} \;, \quad
b_{\hat \theta}= -{n_{F\,\hat\theta} n_{F\,\hat\phi} \over (1-n_{F\,\hat r} )}\;, \quad
b_{\hat \phi}= 1 - {n_{F\,\hat\phi}^2\over(1-n_{F\,\hat r})}\;.
\label{eq:bdef}
\end{eqnarray}
\item
Continue and conclude with steps (ii) and (iii) of \ref{subsec:CircAtCamera}.

\end{enumerate}

\subsubsection{Foundations for these ray-bundle prescriptions}

For completeness we briefly describe the Pineault-Roeder \cite{PR1,PR2} 
foundations for these ray-bundle-evolution prescriptions.

At an arbitrary location $\zeta$ along the reference ray, in the reference frame
of a FIDO there, denote by $\{n_{F\,\hat r}\,, n_{F\, \hat\theta}\,, n_{F\, \hat \phi}\}$
the components of the fiducial ray's unit tangent vector.  (These are the same
$n_F$'s as in the previous two subsections, extended to the fiducial ray's
entire world line.)  And at this arbitrary $\zeta$, introduce the unit basis vectors 
$\bf a$ and $\bf b$ [Eqs.\ (\ref{eq:bdef})] that 
span the plane orthogonal to the fiducial ray.

Notice that far from the black hole (at the celestial sphere), 
where $n_{F\,\hat r}=-1$ (negative because the 
reference ray is incoming), $n_{F\,\hat\theta} = n_{F\,\hat\phi}=0$, the 
only nonzero components of these basis vectors are $a_{\hat \theta} = 1$ and
$b_{\hat \phi}=1$; so ${\bf a} = {\bf e}_{\hat \theta}$ and ${\bf  b} = {\bf  e}_{\hat \phi}$.

We can think of the transverse plane at $\zeta$ as the complex plane, and 
think of the $\bf a$ direction as the real axis for complex numbers, and 
$\bf b$ as the imaginary axis.  Then any transverse vector $\bf Y$ can be
represented by a complex number $Y$, with 
\begin{equation}
{\bf Y} = Y_a \,{\bf a} + Y_b\, {\bf b}\;, \quad Y = Y_a + i Y_b\;.
\label{eq:VecAsComplexNo}
\end{equation}

To describe the ray bundle and its evolution, (following Pineault and Roeder) we introduce two complex numbers
(transverse vectors) $\xi$ and $\eta$, whose real and imaginary parts are functions
that are evolved by the ray-bundle equations (\ref{eq:BundleEvolution}):
\begin{equation}
\xi = u+iv\;, \quad \eta = g+ih\;.
\label{eq:xieta}
\end{equation}
The outer edge of the ray bundle, at location $\zeta$ along the reference ray, 
is described by the complex number 
\begin{equation}
Y = ( \xi \,e^{i \sigma} + \eta\, e^{-i \sigma}) e^{i\chi}\;,
\label{eq:Ydef} 
\end{equation}
where $\chi$ is also evolved by the ray-bundle equations.
Here $\sigma$ is a parameter that ranges from $0$ to $2\pi$, 
and as it changes, it 
carries the head of the vector $\bf Y$ around the bundle's edge. That vector, 
of course, is given by 
\begin{equation}
{\bf Y} = {\Re}[(\xi\, e^{i \sigma} + \eta \,e^{-i\sigma})e^{i\chi}]{\bf a} +
{\Im}[(\xi \,e^{i \sigma} + \eta \,e^{-i\sigma})e^{i\chi}] {\bf b}\;
\label{eq:vecYval}
\end{equation}
[cf.\ Eq.\ (\ref{eq:VecAsComplexNo})], and its tail sits on the reference ray while its
head sits on the edge of the bundle.
One can verify that the shape of this ${\bf Y}(\sigma)$, with $\sigma$ varying
from 0 to $2\pi$, is an ellipse.

Let us denote by $\nu$ and $\lambda$ the arguments of $\xi$ and $\eta$,
so $\xi= |\xi|\, e^{i\nu}$, $\eta = |\eta| e^{i\lambda}$.  Then Eq.\ (\ref{eq:Ydef}) reads
$Y= |\xi| e^{i(\nu+\sigma+\chi)} +  |\eta| e^{i(\lambda-\sigma+\chi)}$.  
As $\sigma$ varies from 0 to $2\pi$,
the modulus of $Y$ (equal to the length of the vector $\bf Y$) reaches its largest value
when the phase of the two terms is the same, i.e.\ when $\nu+\sigma = \lambda - \sigma$,
or equivalently when $\sigma= \frac12(\lambda-\nu)$.  This maximal value of $|Y|$ is
\begin{equation}
|{\bf Y}|_{\rm max} = |Y|_{\rm max} = |\xi|+|\eta|\;,
\end{equation} 
and the argument of $Y$ at the max (the angle of 
$\bf Y$ to $\bf a$) is 
\begin{equation}
\mu = \nu + \sigma +\chi = \frac12(\nu+\lambda) +\chi = 
\frac12 \arg(\xi\, \eta) +\chi\;.
\label{eq:muevolve}
\end{equation}
The major diameter (not angular diameter) of the elliptical ray bundle, as 
measured by a FIDO at the location $\zeta$, is this
$|{\bf Y}|_{\rm max}$ multiplied by $\delta_F = \delta_{\rm cs} (f_c/f_F)$:  
\begin{equation}
\textrm{(major diameter of ellipse)} = \delta_F |{\bf Y}|_{\rm max}
= \delta_F (|\xi|+|\eta|)\;.
\label{eq:majordiameter}
\end{equation}

When the bundle reaches the celestial sphere, its measured
angular diameter is the rate of increase of this major diameter with 
distance traveled.  But at the celestial sphere, the distance traveled is equal to the affine
parameter $\zeta$, so the measured angular diameter is $\delta_F 
(d|{\bf Y}|_{\rm max}/d\zeta)$. Now, the FIDO-measured light frequency 
$f_F$ is equal to $f\,'$ at the celestial sphere, and the
real and imaginary parts of $\xi = u+iv$ and $\eta=g+ih$ are increasing linearly 
with $\zeta$, so it turns out that the formula $\delta_F (d|{\bf Y}|_{\rm max} /
d\zeta)$
becomes the first of Eqs.\ (\ref{eq:deltapm}); and the angle $\mu$ of the major axis 
to the real axis $\bf a$, Eq.\ (\ref{eq:muevolve}), becomes Eq.\ 
(\ref{eq:mu}).  

By an argument similar to the first part of the paragraph before last, one can 
deduce that anywhere along the evolving ray bundle 
\begin{equation}
\textrm{(minor diameter of ellipse)} = \delta_F |{\bf Y}|_{\rm min}
= \delta_F (|\xi|-|\eta|)\;,
\label{eq:majordiameter}
\end{equation}
and correspondingly that the minor angular diameter at the celestial sphere is
the second of Eqs.\ (\ref{eq:deltapm}). 

The evolution equations (\ref{eq:BundleEvolution}) are the equation of geodesic deviation for the
separation vector whose spatial part in a FIDO reference frame is the 
vector ${\bf Y}$ [Eq.\ (\ref{eq:vecYval})] at fixed $\sigma$; see Section II of Pineault and  Roeder \cite{PR2}, and see \cite{KipMathematica}.
\subsection{Filtering}
\label{subsec:App-Filtering}

\subsubsection{Spatial filtering and how we handle point stars}
\label{sec:PointStars}

In DNGR we treat stars as point sources of light with infinitesimal angular size. We trace rays backwards
from the camera to the celestial sphere.  If we were to treat these rays as infinitely thin,
there would be zero probability of any ray intersecting a star; so instead we construct a
beam with a narrow but finite angular width, centred on a pixel on the camera's sky and extending over a small number
of adjacent pixels,  we evolve the shape of this beam along with the ray, and if the beam intercepts the star, we collect the star's light into it.
This gives us some important benefits:

\begin{itemize}
\item The images of our unresolved stars remain small: they don't stretch when they get magnified by
gravitational lensing.

\item The fractional change in the beam's solid angle is directly related to the optical magnification
due to gravitational lensing and hence to the intensity (brightness) change in the image of an unresolved star.

\item When sampling images of accretion discs or extended structures such as interstellar dust clouds,
we minimise moir\'e artefacts by adapting the resampling filter according to the shape of the beam
\cite{GRE86}.
\end{itemize}

Another consequence is that
each star contributes intensity to several pixels. The eye is sensitive to the sum of these  and as a single star crosses this grid, this sum can vary depending on the phase of the geometric image of the star on the grid.\footnote{The same is true, of course, for each bit of a nebula or accretion disk.  We  discuss some details of how DNGR handles accretion disks in \ref{subsec:DNGRdisks}}
This can cause distracting flickering of stars as they traverse the virtual camera's image plane. We mitigate this effect by modulating the star's intensity across the beam with a truncated Gaussian filter and by setting the beam's initial radius to twice the pixel separation.

With this filter, the sum is brightest when the geometric image of the star falls exactly on a pixel and dimmest when centred between four pixels. The shape and width of the filter are designed to give a maximum 2\% difference between these extremes, which we  found, empirically, was not noticeable in fast changing \emph{Interstellar} scenes.

This result assumes the final size of the ellipse is small and the shape of the beam does not change significantly between adjacent pixels. In extreme cases these assumptions can break
down, leading to a distortion in the shape of a star's image, flickering, and aliasing artefacts. 
In these cases we can trace multiple beams per pixel and resample (See e.g. Chapter 7 of \cite{PBRT}).
However, \emph{Interstellar}'s black hole, Gargantua, when visualized,
has a moderate
spin $a/M=0.6$, which gives rise to few areas of extreme distortion, so we did not observe these problems in images created for the movie.

\subsubsection{Temporal filtering: motion blur}
For moving images, we need to filter over time as well as space. A traditional film camera typically exposes the film for half the time between individual frames of film, so for a typical 24 fps (frame per second) film, the exposure time will be 1/48 s. (This fraction is directly related to the camera's ``shutter angle'': the shutters in film cameras are rotating discs with a slice cut out. The film is exposed when the slice is over the emulsion, and the film moves onto the next frame when covered. The most common shape is a semicircular disk which is a ``180 degree shutter''.) Any movement of the object or camera during the exposure leads to motion blur. In our case, if the camera is orbiting a black hole, stellar images close to a critical curve appear to zip around in frantic arcs and would appear as streaks of light on a 1/48 s photograph. We aim to reproduce that effect when creating synthetic images for films.

This motion blur in computer graphics is typically simulated with Monte Carlo methods, computing multiple rays per pixel over the shutter duration. In order to cleanly depict these streaks with Monte Carlo methods, we would need to compute the paths of many additional rays and the computational cost of calculating each ray is very high. Instead, in DNGR we take an analytic approach to motion blur (cf.\ Figure \ref{FigA1:MotionBlur})  by calculating how the motion of the camera influences the motion of the deflected beam:

\begin{figure}[t!]
\begin{center}
\includegraphics[width=1.0\columnwidth]{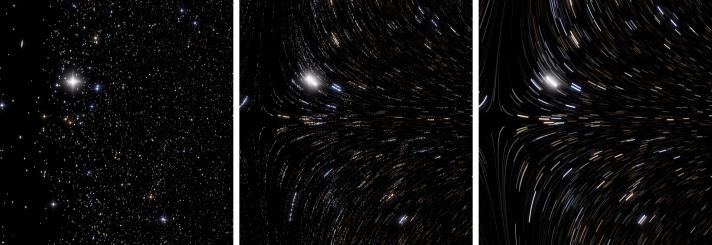}
\end{center}
\caption{Left-to-right: No motion blur; Monte Carlo motion blur with four time samples per-pixel; Analytic motion blur. Relative computation time is approximately in the ratio 1:4:2. Lens flare has been added to each image
to illustrate how these images would typically be seen in the context of a movie.}
\label{FigA1:MotionBlur}
\end{figure}

The camera's instantaneous position influences the beam's momentary state, and likewise the camera's motion affects the time derivatives of the beam's state. We augment our ray and ray bundle equations (\ref{eq:Rays2}) and (\ref{eq:BundleEvolution}) to track these derivatives throughout the beam's trajectory and end up with a description of an elliptical beam being swept through space during the exposure. The beam takes the form of a swept ellipse when it reaches  the celestial sphere, and we integrate the contributions of all stars within this to create the motion-blurred 
image.\footnote{We took a similar approach when motion-blurring the accretion disks discussed in \ref{subsec:DNGRdisks}.} These additional calculations approximately double the computation time, but this is considerably faster than a naive Monte Carlo implementation of comparable quality.

\subsubsection{Formal description of our analytic approach to motion blur}

We can put our analytic method into a more formal setting, using the mathematical description of motion blurred
rendering of Sung {\it et al.} \cite{SUNG}, as follows.  We introduce the quantity
\begin{equation}
I_{xy} = \sum _l\int_{\Omega}\! \int_{\Delta T} r(\omega, t) g_l(\omega, t) L_l(\omega, t) \,\mathrm{d}t
\,\mathrm{d}\omega\;,
\end{equation}
which represents the resulting intensity of a sample located at coordinates $(x,y)$ on the virtual image plane of the camera, corresponding to a ray direction $\omega$, and subtending a solid angle, $\Omega$. The sum is over each object, $l$, in the scene, $\Delta T$ is the time the shutter is open, and $L_l(\omega, t)$ represents the radiance of the object directed back towards the camera along $\omega$ from object $l$. The shutter and optics may filter the incident radiance (specific intensity) and this is accounted for by the term $r(\omega, t)$. The term $g_l(\omega, t)$ is a geometric term that takes the value 1 if object $l$ is visible or 0 if occluded.

Considering just the distortion of the celestial sphere to start with, we start with a ray in the direction $\omega_0$ at time $t_0$, in the middle of the exposure. We determine whether this ray gets swallowed up by the black hole or continues to the celestial sphere using equations (\ref{eq:b1b2}) and (\ref{eq:qob}). This corresponds to $g(\omega_0, t_0)$.
We evolve the initial direction $\omega$ and solid angle $\Omega$ subtended by the sample into a new direction at the celestial sphere, $\omega^\prime$, and ellipse $\Omega^\prime$, using the method described in  \ref{subsec:App-RayTrace}
and \ref{subsec:App-RayBundle}.

When the ray strikes the celestial sphere, we integrate over the area $\Omega^\prime$, weighting the integration with our truncated Gaussian filter, corresponding to $r(\omega)$. (We assume there is no
time-dependent aspect to this filter and ignore the $t$ term. We also assume the celestial sphere
is static.)

Our motion blur method approximates $L(\omega)$ with the following expression:
\begin{equation}
L\left(\omega^\prime + t\left(\frac{\partial \omega^\prime}{\partial t} \right)_{\!\omega_0, t_0} \right)\;,
\end{equation}
where the partial derivative is evaluated at the centre of the beam $\omega_0$ and at $t_0$.  This represents a ray that can be swept over an arc in the celestial sphere by varying the value of $t$. This is convolved with the beam shape $\Omega^\prime$ and filter and occlusion terms to give:
\begin{equation}
I_{xy} = g(\omega_0, t_0)\int_{\Omega^\prime} \!\int_{\Delta T} r(\omega) 
L\left(\omega^\prime + t\left(\frac{\partial \omega^\prime}{\partial t} \right)_{\!\omega_0, t_0} \right) \,\mathrm{d}t
\,\mathrm{d}\omega^\prime\;.
\end{equation}

In this form, the consequences of our approximations become clearer:
By evaluating $g$ only at the beam centre $\omega_0$ and shutter-centre $t_0$, a sample that crosses the black hole's shadow  during
the exposure would be given a constant value of $g$ (either 0 or 1) instead of the correct expression that transitions during the exposure. This would lead to a hard edge at the shadow if the camera performed a fast pan.

Luckily, in \emph{Interstellar}, the design of the shots meant this rarely happened: the main source of motion
blur was the movement of the spaceship around the black hole, and not local camera motion.

To handle any cases where this approximation might cause a problem, we also implemented a hybrid method where we
launch a small number of Monte Carlo rays over the shutter duration and each ray sweeps a
correspondingly shorter analytic path.

We used a very similar technique for the accretion disk.

\subsection{Implementation}
\label{subsec:App-Implementation}

DNGR was written in C++ as a command-line application. It takes as input the camera's position, velocity, and field of view, as well as the black hole's location, mass and spin, plus details of any accretion disk,  star maps and nebulae.

Each of the 23 million pixels in an IMAX image defines a beam that is evolved as described in \ref{subsec:App-RayBundle}.

The beam's evolution is described by the set of 
coupled first and second order differential equations (\ref{eq:Rays2}) and
(\ref{eq:BundleEvolution}) that we put into fully first order
form and then numerically integrate  backwards in time using a custom implementation of the Runge-Kutta-Fehlberg method (see, e.g., Chapter 7 of \cite{RKF}).  This method gives an estimate of the truncation error at every integration step so we can adapt the step size during integration: we take small steps when the beam is bending sharply relative to our coordinates, and large steps when it is bending least. We use empirically determined tolerances to control this behaviour. 
Evolving the beam along with its central ray triples the time per integration step, on average,
compared to evolving only the central ray.

The beam either travels near the black hole and then heads off to the celestial sphere; or it goes into the black hole and is never seen again; or it strikes the accretion disk, in which case it gets attenuated by the disk's optical thickness  to extinction or continues through and  beyond the disk to pick up additional data from other light sources, but with attenuated amplitude. 
We use automatic differentiation \cite{Piponi}  to track the derivatives of the camera motion
through the ray equations.

Each pixel  can be calculated independently, so we run the calculations in parallel over multiple CPU cores and over multiple computers on our render-farm.

We use the OpenVDB library \cite{OPENVDB} to store and navigate volumetric data and Autodesk's Maya \cite{MAYA} to design the motion of the camera. (The motion is chosen to fit the film's narrative.)  A custom plug-in running within Maya creates the command line parameters for each frame. These commands are queued up on our render-farm for off-line processing.

\subsection{DNGR Code characteristics and the Double Negative render-farm}
\label{subsec:App-CodeFarm}

A typical IMAX image has 23 million pixels, and for {\it Interstellar} we had to generate many thousand images, so DNGR had to be very efficient.  It has 40,000 lines of C++ code and runs across Double Negative's Linux-based render-farm. Depending on the degree of gravitational lensing in an image,\footnote{More specifically: the render time depends mainly on the number of steps in the integration.  This, in turn, depends on the truncation errors in the integration scheme.  The tolerances are tighter on the position of the ray than its shape, as errors in position give rise to more noticeable artefacts.  Empirically, renders 
of regions close to the black hole's shadow
are much slower than any other.}
it typically takes from 30 minutes to 
several hours running on 10 CPU cores to create a single IMAX image.  The longest renders were those of the close-up accretion disk when we shoe-horned DNGR into Mantra.  For \emph{Interstellar}, render times were never a serious enough issue to influence shot composition or action.

Our London render-farm comprises 1633 Dell-M620 blade servers; each blade has two 10-core E5-2680 Intel Xeon CPUs with 156GB RAM.
During production of {\it Interstellar}, several hundred of these were typically being used by our DNGR code.

\subsection{DNGR modelling of accretion disks}
\label{subsec:DNGRdisks}

Our code DNGR includes three different implementations of an accretion disk:

\emph{Thin disk}:
For this, we adapted our DNGR ray-bundle code so it detects intersections of the beam with  the disk.  At each intersection,  we sample
the artist's image 
in the manner described in Section \ref{subsec:RayBundle} and
\ref{subsec:App-Filtering} above,
to determine the colour and intensity of the bundle's light and we attenuate the beam  by the optical thickness (cf.\ \ref{subsec:App-Implementation}).  

\emph{Volumetric disk}:
The volumetric accretion disk was built by an artist using SideFX Houdini software and stored in a volumetric data structure containing roughly 17 million voxels (a
fairly typical number). Each voxel describes the optical density and colour of the disk in that region. 

We used Extinction-Based Sampling \cite{KRAUS}
to build a mipmap volume representation of this data.
The length of the ray bundle was split into short piecewise-linear segments which, in turn, were traced through the voxels.
We used the length of the major axis of the beam's cross section to select the closest two levels in the mipmap volume; and in
each level we  sampled the volume data at steps the length of a voxel, picking up contributions from the colour at each voxel
and attenuating the beam by the optical thickness.  
The results from the two mipmap levels were interpolated before
moving on to the next line segment.

\emph{Close-up disk with procedural textures}:
Side Effects Software's renderer, Mantra, has a plug-in architecture that lets you modify its operation. We embedded the DNGR ray-tracing code into a plug-in and used it to generate piecewise-linear ray segments which were evaluated through Mantra. This let us take advantage of Mantra's procedural textures\footnote{A \emph{procedural texture}  uses an algorithm, such as fractal noise, to generate detail at arbitrary resolution, unlike a texture based on an image file, which has a finite resolution.} and shading language to create a model of the accretion disk with much more detail than was possible with the limited resolution of a voxelised representation. However this method was much slower so was only used when absolutely necessary.

\vskip1pc

\emph{Disk layers close to the camera}   
The  
accretion-disk images in \emph{Interstellar} were generated in
layers that were blended together to form the final images. Occasionally
a layer only occupied space in the immediate vicinity of the camera, so close that
the infuences of spacetime curvature and gravitational redshifts were negligible.
These nearby layers were 
rendered as if they were in flat spacetime, to reduce computation time.

\emph{Frequency shifts.} 
In our  visualizations that included Doppler and gravitational frequency shifts, we modelled the accretion disk as a constant-temperature blackbody
emitter and we estimated the star temperatures from published NASA data \cite{Tycho2}. The Doppler and gravitational frequency shifts correspond to
a temperature shift in the blackbody spectra on arrival at the camera. We convolve those temperature-shifted black body spectra with the published spectral sensitivity curves of a
typical motion picture film \cite{KODAK} to generate separate red, green and blue radiance values, $R$, $G$ and $B$.
For the volumetric disk, these are calculated at each step and used in place of
 artist-created radiance values. 
In Figure \ref{Fig15:Disk}b we removed the doppler-induced intensity change by dividing Figure \ref{Fig15:Disk}d's RGB triplet
of radiance values by a weighted mean that, empirically, leaves the eye-perceived colours unchanged:
$\{R^\prime, G^\prime, B^\prime\} = \{R,G,B\}/(0.30 R + 0.59 G + 0.11 B)$.  This dimmed the blue side of the disk and brightened the red side.   

We set the white balance of our virtual camera to render a 6500 K blackbody spectrum
with equal red, green and blue pixel values by applying a simple gain to each colour channel.
We did not model the complex, nonlinear interaction between the colour-sensitive layers that
occurs in real film.

\subsection{Comparison of DNGR with astrophysical codes and with film-industry CGI codes}
\label{App:CompareCodes}

Near the end of Section \ref{subsec:Previous}, we compared our code DNGR with the state-of-the-art astrophysical visualization code GRay.  The most important differences---DNGR's use of light-beam
mappings vs.\ GRay's use of individual-ray mappings, and DNGR's use of ordinary processors vs.\ GRay's use of GPUs---were motivated by our different goals: smoothness of images
in movies vs.\ fast throughput in astrophysics.

So far as we are aware, our code is unique in using light-beam-based mappings.  No other code, astrophysical or film CGI, uses them.  However, some film CGI codes use flat-spacetime mappings from the camera's image plane to the source plane
that are mathematically equivalent to our light beams.  Specifically, they use \emph{ray differential} techniques described by Igehy \cite{Igehy99}, which track, along a ray, the transverse derivatives of the locations of nearby rays, thereby producing the matrix
$||\partial \xi^i/\partial x^j||$, where $\xi^i(x^j)$ is a ray's source-plane location as a function of its image-plane location.  Tracking these differentials in flat spacetime is 
moderately easy because they are constant along straight paths between reflections,
refractions, etc; and they change via simple matrix transformation laws at interaction locations.  Tracking them through curved spacetime would entail the same 
geodesic-deviation technique as underlies our ray-bundle mappings.

In  \ref{sec:PointStars} we have described our methods of imaging star fields, using stars that are point sources of light:  we feed each star's light into every light beam that intersects the star's location, with an appropriate weighting.

Other gravitational-lensing codes deal with star fields differently.  For example, 
M\"uller and Frauendiener \cite{Muller12} calculate where each point-star ends up in the camera's image plane, effectively producing the
inverse of the traditional ray-tracing algorithm. They do this for lensing by a non-spinning
black hole (Schwarzchild metric) which has a high degree of symmetry, making this calculation tractable.
Doing so  for a spinning black hole would be much more challenging.

A common method for rendering a star-field is to create a 2D source picture (environment map) that records
the stars' positions as finite-sized dots in the source plane (see e.g. \cite{BOURKE}).
This source picture is then sampled using the evolved rays from the camera. This has the disadvantage that 
stars can get stretched in an unrealistic way in areas of extreme magnification, such as near the
critical curves described in Section \ref{subsec:KerrCaustics} of this paper.  As we discussed in  \ref{sec:PointStars}, our DNGR light-beam technique with point stars circumvents this  problem---as does the M\"uller-Frauendiener technique with point stars.

\end{document}